\documentclass[%
reprint,
superscriptaddress,
amsmath,amssymb,
aps,
prb,
]{revtex4-1}

\usepackage{graphicx}
\usepackage{dcolumn}
\usepackage{bm}
\usepackage{mathtools}
\usepackage{bbold}
\usepackage{color}

\begin{document}

\preprint{}

\title
{Critical Currents in Conventional Josephson Junctions With Grain Boundaries}

\author{Miguel Antonio Sulangi}
\affiliation{
	Department of Physics, University of Florida, Gainesville, FL 32611, USA
}

\author{Laetitia Bettmann}
\affiliation{
	Institut f{\"u}r Theoretische Physik, Universit{\"a}t Leipzig, D-04103 Leipzig, Germany
}

\author{T.A. Weingartner}
\affiliation{Department of Electrical and Computer Engineering,
 University of Florida, Gainesville, FL 32611
}
\author{N. Pokhrel}
\affiliation{Department of Electrical and Computer Engineering,
 University of Florida, Gainesville, FL 32611
}
\author{E. Patrick}

\affiliation{Department of Electrical and Computer Engineering,
 University of Florida, Gainesville, FL 32611
}
\author{M. Law}
\affiliation{Department of Electrical and Computer Engineering,
 University of Florida, Gainesville, FL 32611
}

\author{A. Kreisel}
\affiliation{
	Institut f{\"u}r Theoretische Physik, Universit{\"a}t Leipzig, D-04103 Leipzig, Germany
}

\author{P. J. Hirschfeld}
\affiliation{
	Department of Physics, University of Florida, Gainesville, FL 32611, USA
}

\date{\today}

\begin{abstract}
It has been hypothesized that the variation of the critical currents in Nb/Al-AlO$_x$/Nb junctions is due to, among other effects, the presence of grain boundaries in the system. Motivated by this, we examine the effect of grain boundaries on the critical current of a Josephson junction. We assume that the hopping amplitudes are dependent on the interatomic distance and derive a physically realistic model of distance-dependent hopping amplitudes. We find that the presence of a grain boundary and associated disorder is responsible for a very large drop in the critical current relative to a clean system. We also find that when a tunnel barrier is present, grain boundaries cause substantial variations in the critical currents due to the disordered hoppings near the tunnel barrier. We discuss the applicability of these results to Josephson junctions presently  intended for use in superconducting electronics applications.
\end{abstract}

\maketitle

\section{\label{sec:level1}Introduction}

Considerable progress in the field of superconducting electronics has recently been made possible by advances in the fabrication of superconducting devices.\cite{McCaughan2014,tolpygo2016superconductor} A key difficulty, however, is that, for many candidate superconducting devices intended for use in applications, such as the Nb/Al-AlO$_x$/Nb junctions\cite{gurvitch1983high} currently being developed by the MIT Lincoln Laboratory, there is considerable device-to-device variability in the Josephson critical current.\cite{tolpygo2014fabrication, holmes2017non, tolpygo2017properties,tolpygo2019advanced} This is a stumbling block for the practical applications of these devices, which frequently involve a very large number of Josephson junctions; an important requirement is that the variations of all device parameters are accounted for.  

Many proposals have been put forth to explain this device-to-device variability. These include thickness variations in the oxide (tunnel) barrier layer, the presence of pinholes (\emph{i.e.}, portions of the Josephson junction where the tunnel barrier is absent), and vacancies.\cite{tolpygo2014fabrication} These scenarios were explored in detail in our previous work, which demonstrated how realistic amounts of device-to-device variability can naturally result from the presence of these forms of disorder.\cite{sulangi2020disorder} Our current work demonstrates a physically plausible mechanism for critical-current variability distinct from previous explanations, related to the presence of grain boundaries within the superconducting niobium leads. This has been proposed as a possible explanation for the aforementioned critical current variation ; more importantly, recent device-simulation work has demonstrated that grain boundaries are a natural consequence of the processes employed in the fabrication of these Josephson junctions\cite{pokhrel2021modeling}. Grain boundaries have been shown to lead to anisotropy in the electromechanical properties of bulk superconductors.\cite{santhanam1976flux, dasgupta1978flux, zerweck1981pinning, garg2020possible} For the case of Al/AlO$_x$/AlO junctions, it has also been argued that grain boundaries lead to variations of the thickness of the oxide barrier.\cite{nik2016correlation} What is not immediately clear, however, is the precise way in which these grain boundaries affect the critical current of these Josephson junctions.

Much guidance can be gleaned, however, from earlier work on grain boundaries in the cuprate high-temperature superconductors. Here, it was shown that grain boundaries cause a sharp suppression of the critical current, with a roughly exponential dependence on the misorientation angle.\cite{dimos1988orientation, hilgenkamp2002grain} Theoretical and numerical simulations have reproduced this critical-current suppression using a model of grain boundaries in $d$-wave superconductors.\cite{graser2010grain, wolf2012supercurrent} It was found that near the grain boundary, the interatomic hopping amplitudes are highly disordered due to the presence of atoms whose positions deviate from that of a perfectly ordered lattice, and this, along with the charge buildup arising from many species of atoms present in the cuprates, accounts for the suppression of the critical current. In an $s$-wave superconductor such as Nb, much less sensitive to disorder, 
the effect of grain boundaries is expected to be quite different, however.

In this work, we use a multi-step approach to treating the problem of a grain boundary located within an $s$-wave niobium superconductor,  following the prescription in Ref. \onlinecite{graser2010grain}. We first perform molecular dynamics simulations to determine the positions of the niobium atoms. Next, we obtain, within a  simplified model, the distance-dependence of the hopping amplitudes by calculating the overlap between orbitals belonging to two spatially separated atoms. Finally, using the atomic positions and the distance-dependence of the hopping amplitudes as input information, we calculate the critical current of this junction from the exact diagonalization of the mean-field Bogoliubov-de Gennes Hamiltonian, after imposing a phase gradient across the system and iteratively solving the system until self-consistency (or equivalently, current conservation\cite{furusaki1991dc, sols1994crossover}) is reached.

We find that for a junction consisting solely of a grain boundary oriented perpendicularly to the phase gradient, the critical-current density is highly suppressed relative to the intrinsic critical current of a clean bulk superconductor. The critical-current suppression is found to be large, often around half of the clean limit. We find that this suppression is due to the disordered hopping amplitudes near the grain boundary, which cause a sharp phase drop in the vicinity of the grain boundary and, therefore, act to decrease the smooth phase gradient away from the grain boundary, resulting in the reduction of the critical-current density. We find that the critical current is only weakly dependent on the misorientation angle, and that various types of grain boundaries cause critical-current suppressions of roughly similar orders of magnitude.

We also examine the case where an oxide tunnel barrier is present together with various grain boundaries. We find that the presence of grain boundaries generically causes variations in the resulting critical current, which depend on the configuration of the grains in the system. This is a consequence of the highly disordered positions of the atoms near the tunnel barrier, which cause hopping amplitudes to exhibit large variations in magnitude and, therefore, lead to enhanced or suppressed transmission through various portions of the barrier depending on the hopping strength of the nearby atoms. We find that the critical current can even be enhanced due to the presence of these grain boundaries.

\section{Model and Methods} \label{model}

We employ a multi-step process to model the effects of a grain boundary on the critical current of a superconducting system. First, molecular dynamics simulations are performed to obtain the positions of the atoms in the presence of a grain boundary. We then calculate the hopping amplitudes as a function of interatomic distance from a model of orbital overlaps between two atoms, which are spatially separated. Finally, we  set up a Hamiltonian in real space using this information and we then calculate the critical current after obtaining self-consistent mean-field solutions using the Bogoliubov-de Gennes approach.\cite{blonder1982transition,furusaki1991dc,furusaki1994dc,sols1994crossover,asano2001numerical,nikolic2002equilibrium,andersen2005bound,andersen20060,covaci2006proximity,andersen2008josephson,black2008self,graser2010grain,wolf2012supercurrent,sulangi2020disorder} We outline these steps in detail in the following subsections.

\subsection{Molecular Dynamics Modeling of Atomic Positions}
\begin{figure}[tb]
	\centering
\includegraphics[width=\linewidth]{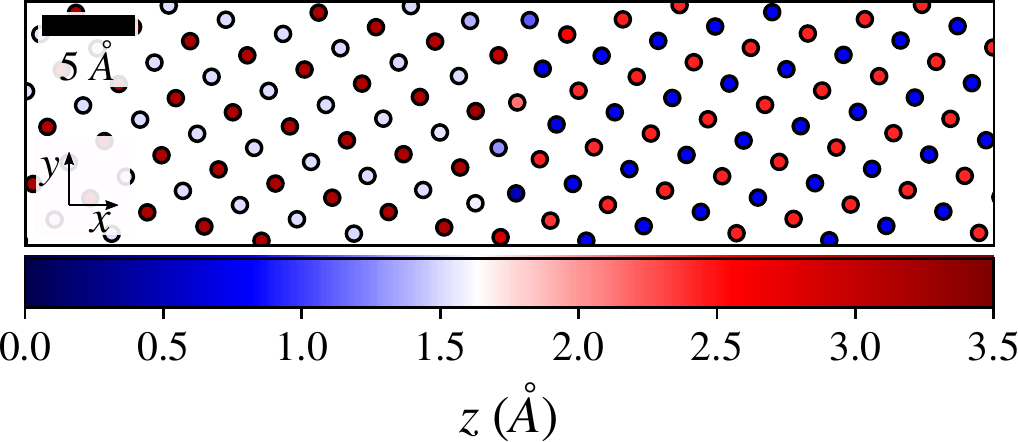}
	\caption{Top view of the atomic positions of the  symmetric (410) GB. The ``red'' and ``blue'' sub-lattices are the two-dimensional systems being considered in the self-consistent BdG calculations. The atomic $z$-positions are included in calculating the hopping through the  magnitude of the distance vector. Throughout, the axes in the plane are chosen as indicated by the arrows. The length scale is represented via the scale bar.}
	\label{fig:410z} 
\end{figure}

The equilibrium symmetric tilt grain-boundary (GB) structures (Fig.~\ref{fig:410z}) were created in Large-scale atomic/molecular massively parallel simulator (LAMMPS)\cite{Plimpton1995}. Similar grain boundaries have been simulated recently \cite{Garg2020}. A force-matched embedded-atom method (EAM) potential for niobium was used\cite{Fellinger2010}. The GB interiors in the bicrystalline simulation cell with 3D periodic boundary conditions were chosen sufficiently large to obtain minimum energy GB structures \cite{Rittner1996,Rajagopalan2017,Adlakha2014}. The energy of Nb atom configurations was then minimized using a non-linear conjugate gradient energy method\cite{Rajagopalan2014,Solanki2013,Tschopp2012}, which yields a lattice constant $a_0$ = 3.3079 \AA{} for an undisturbed system, in agreement with the lattice constant of niobium. Here, we have allowed the simulation box to expand or compress perpendicular to the grain-boundary direction. The resulting Nb structures are bcc lattices within the grains. We consider only a simple bilayer (Fig.~\ref{fig:410z}) that is sufficient to model the whole system.\\

\subsection{Calculation of Hopping Amplitudes}
\begin{figure}[tb]
	\centering
\includegraphics[width=\linewidth]{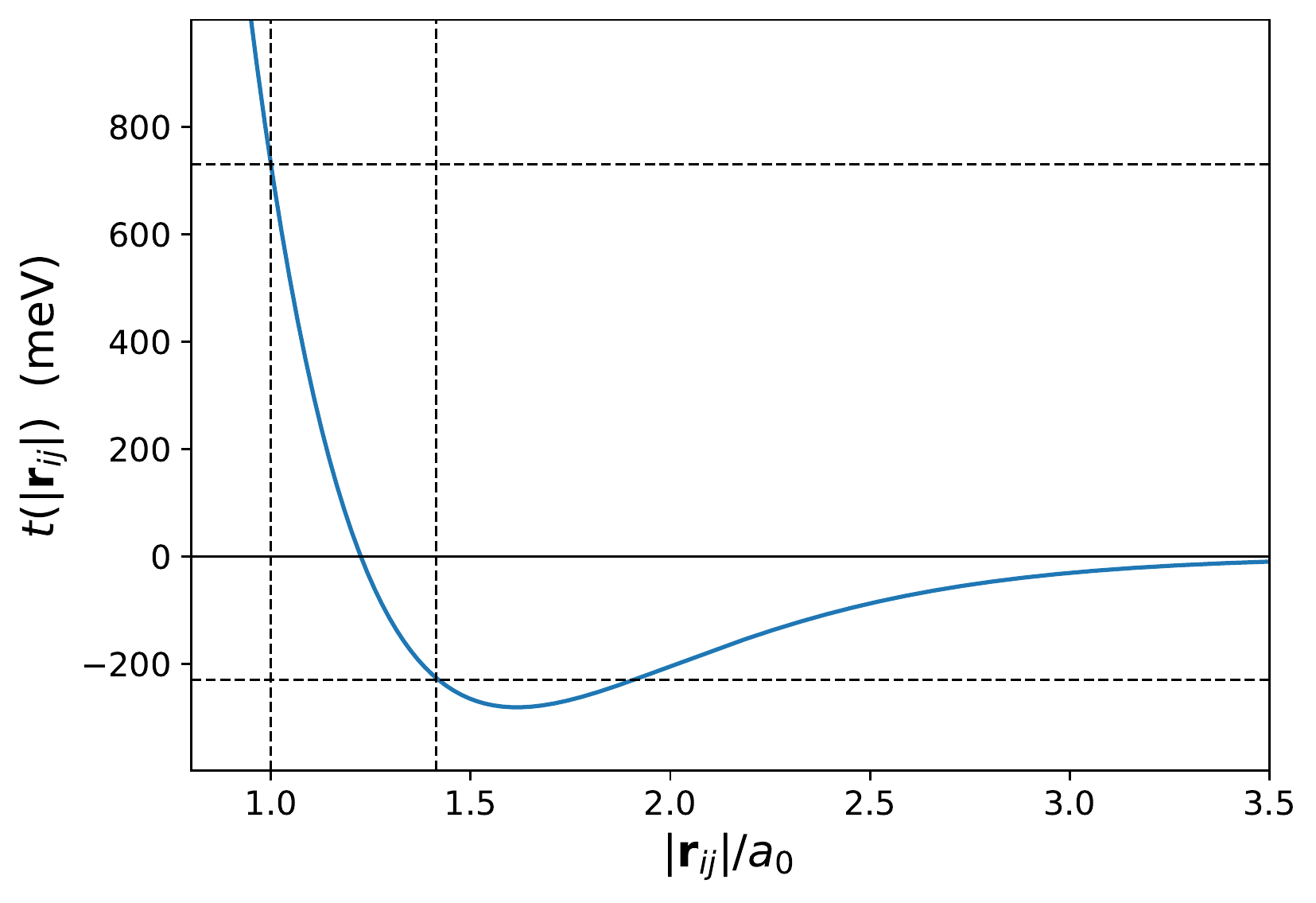}
	\caption{Hopping amplitude $t$ as a function of distance $|\mathbf{r}_{ij}|$ between sites computed from an expectation value of the kinetic energy from overlapping atomic $s$-orbitals. The relevant values at the NN $|\mathbf{r}_{ij}|=a_0$ and NNN $|\mathbf{r}_{ij}|=\sqrt{2} a_0$ distances of the homogeneous system are indicated by dashed lines.}
	\label{fig:hopps} 
\end{figure}

In a lattice distorted in the presence of a GB, the distinction among nearest-neighbor (NN), next-nearest-neighbor (NNN), and next-next-nearest-neighbor (NNNN) hoppings is blurred, and in order to set up a tight-binding model for the system with disorder, one needs to obtain hoppings as continuous functions of the distance between sites $\mathbf{r}_{ij} = \mathbf{r}_i-\mathbf{r}_j$ , i.e., as $t(|\mathbf{r}_{ij}|)$. To connect to earlier investigations\cite{sulangi2020disorder} based on a single-band model, we obtain the function $t(|\mathbf{r}_{ij}|)$ by computing the expectation value of the kinetic energy $-\nabla^2/(2m^*)$ from overlapping atomic $s$-orbitals (see Fig.~\ref{fig:hopps}). This approximation is very simplified and ignores that the electronic structure has significant d-orbital weight; still, it captures the effect that the hopping strongly increases in magnitude if atoms move closer to each other and a much less pronounced weakening of the hoppings as these move further apart; this nonlinear behavior is expected in the real system and will strongly modulate the critical current. To match the physical properties of the Nb system, we tune the effective mass $m^*$ such that the NN and NNN hoppings take the values $t_\text{NN}=730$ meV and $t_\text{NNN}=-230$ meV, which exhibit a maximal Fermi velocity of $1\times 10^6 \mathrm{m/s}$, in rough agreement with the magnitude and degree of anisotropy of the Fermi velocity in Nb. Experimentally, the Fermi velocity has been deduced from de Haas-van Alphen measurements to be between $1.3\times 10^5 \mathrm{m/s}$ and $5.4\times 10^5 \mathrm{m/s}$\;\cite{Bruin2013} which is slightly smaller than a result from an \emph{ab initio} calculation giving the range between $2\times 10^5 \mathrm{m/s}$ and $12\times 10^5 \mathrm{m/s}$ \;\nocite{KoepernikFPLO}\footnote{Obtained from a DFT calculation using the lattice constant of 2.87565467\AA{} in the primitive setting and local orbital basis in the LDSA approximation using the code based on Ref. 40; Y. Quan, private communication.}. The use of an effective one-band model neglects the effects of the  anisotropic multiple Fermi-surface sheets present in actual Nb, which are known to drive a small but significant gap anisotropy, of order 10\%\cite{Hahn98}.

\subsection{Calculation of Critical Current}
To calculate the critical current, we follow the approach taken by Graser \emph{et al.}\cite{graser2010grain} The tight-binding Hamiltonian describing electrons hopping on atomic sites and on-site ($s$-wave) pairing is  the standard lattice BCS Hamiltonian,
\begin{equation}
H =-\sum_{ ij\sigma }  t_{ij}c_{i\sigma}^{\dagger}c_{j\sigma} + \sum_{i}\big(\Delta_{i}c_{i \uparrow}^{\dagger}c_{i \downarrow}^{\dagger} + \text{h.c.}\big).
 \label{eq:hamiltonian}
\end{equation}
Here, $t_{ij}$ is the hopping amplitude between sites $i$ and $j$, which is calculated by evaluating the distance between sites $i$ and $j$, with open boundary conditions imposed along the $x$-direction. $\Delta_{i}$ is the on-site $s$-wave order parameter which is determined self-consistently.  By performing a particle-hole transformation, this Hamiltonian can be recast in Bogoliubov-de Gennes (BdG) form. By diagonalizing the BdG Hamiltonian, we can obtain the order parameter $\Delta_{i}$ on the $i$th site, which is given by

\begin{equation}
\Delta_{i} =\frac{U}{2}\sum_n \Big[u_{n,i} v^*_{n,i} \tanh{\Big(\frac{E_n}{2k_B T}\Big)}\Big].
 \label{eq:orderparameter}
\end{equation}
Here, $u_{n,i}$ and $v_{n,i}$ are the particle and hole amplitudes, respectively, on site $i$ corresponding to the $n$th eigenvector of the BdG Hamiltonian and $U$ is the pairing interaction, which is set to $U = 2$ (in units where $t_\text{NN}$, the nearest-neighbor hopping for an ordered square lattice, is 1).
This choice of pairing interaction gives rise to a gap and critical current in the clean limit that are much bigger than the respective values for real-world niobium. We are forced to use a bigger gap than found experimentally because of the necessarily small sizes of our real-space simulations, which restrict the energy resolution available and make it  difficult for the effects of a very small superconducting gap to be detected.
Finally, the bond current $j_{ij}$ between sites $i$ and $j$ is given by 
\begin{equation}
\frac{j_{ij}}{ e} =-2i t_{ij} \sum_n \big[u_{n,i} u^*_{n,j}f(E_n) +  v_{n,i} v^*_{n,j}f(-E_n) - \text{h.c.}\big].
 \label{eq:bondcurrent}
\end{equation}
In our calculations, we set the temperature to $T = 0.0001$, which is  effectively in the $T \to 0$ limit since $T$ is smaller than any other energy scale. We first impose a phase gradient along a portion of the system between  $x = -d$ and $x = d$. We iteratively solve Eqs.~\ref{eq:hamiltonian} and \ref{eq:orderparameter} until self-consistency is reached, and let the phase of the order parameter evolve freely from Eq.~\ref{eq:orderparameter} except at the portions of the system where $x < -d$ and $x > d$, which are held fixed (for instance, at 0 and $\pi/2$, respectively). Here, our definition of self-consistency is current conservation.\cite{furusaki1991dc, sols1994crossover} We obtain the total current across a number of cross sections spread equidistantly within $-d \leq x \leq d$ using Eq.~\ref{eq:bondcurrent}, and we determine that current conservation is present if all total currents corresponding to different cross sections are within a predefined very small tolerance equal to each other. Open boundary conditions are imposed along the $x$-direction, while we use periodic boundary conditions in the $y$-direction.

\section{Grain-Boundary Junctions}

\begin{figure*}[ht]
	\centering
 	\includegraphics[width=1.0\linewidth]{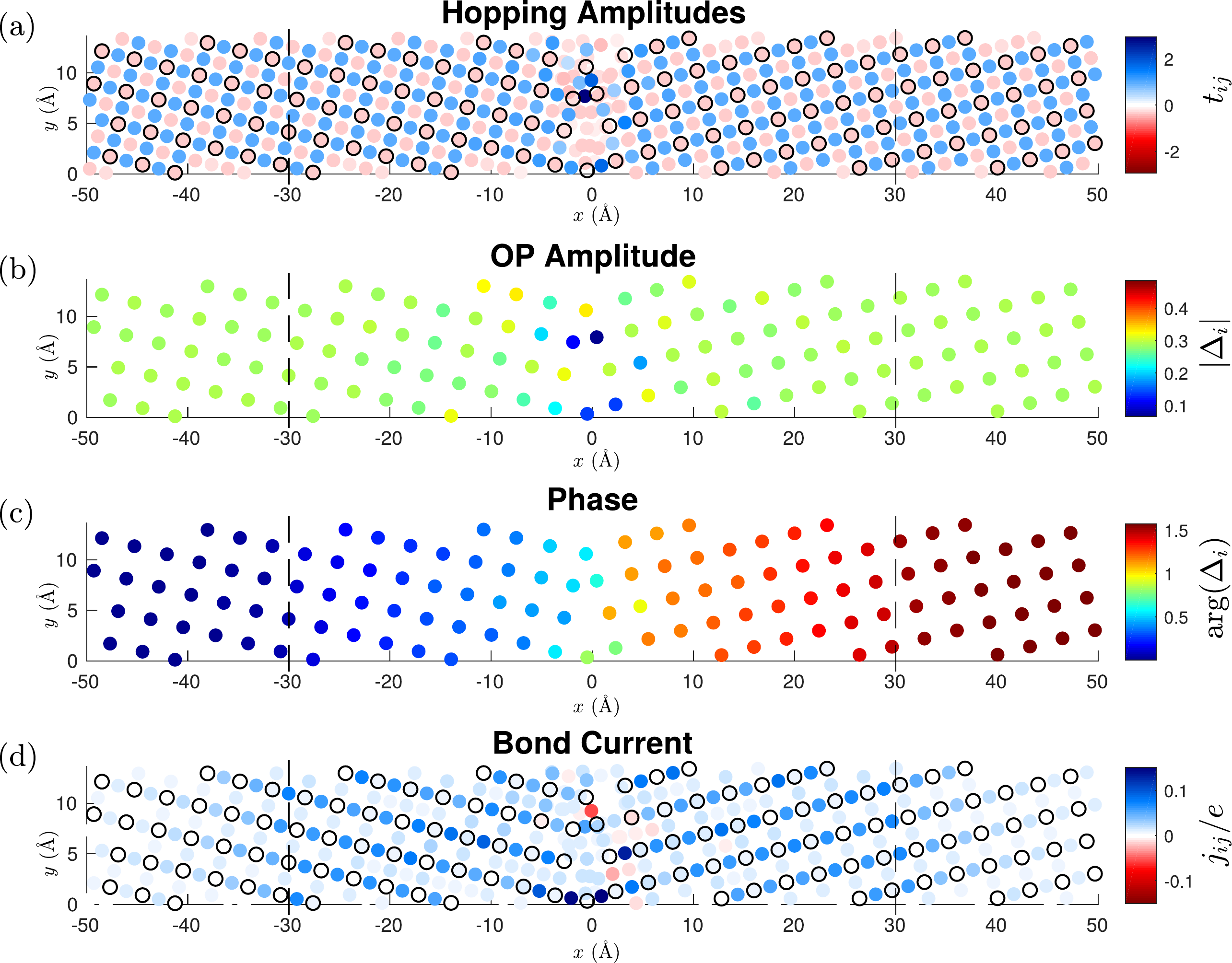}
	\caption{A symmetric (410) grain-boundary junction. (a) Hopping amplitudes as present in the (normal state) Hamiltonian, (b) order parameter amplitude $|\Delta_i|$, (c) order parameter phase $\mathrm{arg}(\Delta_i)$, and (d) bond currents $j_{ij}/e$ in the positive $x$-direction for the self-consistent solution. The quantities are shown for the portion of the system where $-50$ \r{A} $< x < 50$ \r{A}. Here, the phase is held fixed at $x = -30$ \r{A} and $x = 30$ \r{A} (dashed lines in all four plots). For (a) and (d), the hopping amplitudes and bond currents are plotted at the center-of-mass position of any two pairs of atoms, each of which is indicated by an open black circle. Note that the interior color of the circles in (a) and (d), therefore, represents third-nearest-neighbor hopping and bond current, respectively, whose center-of-mass locations coincide with the atomic site locations.}

	\label{fig:410gb}
\end{figure*}

\begin{figure}[ht]
	\centering
	\includegraphics[width=0.5\textwidth]{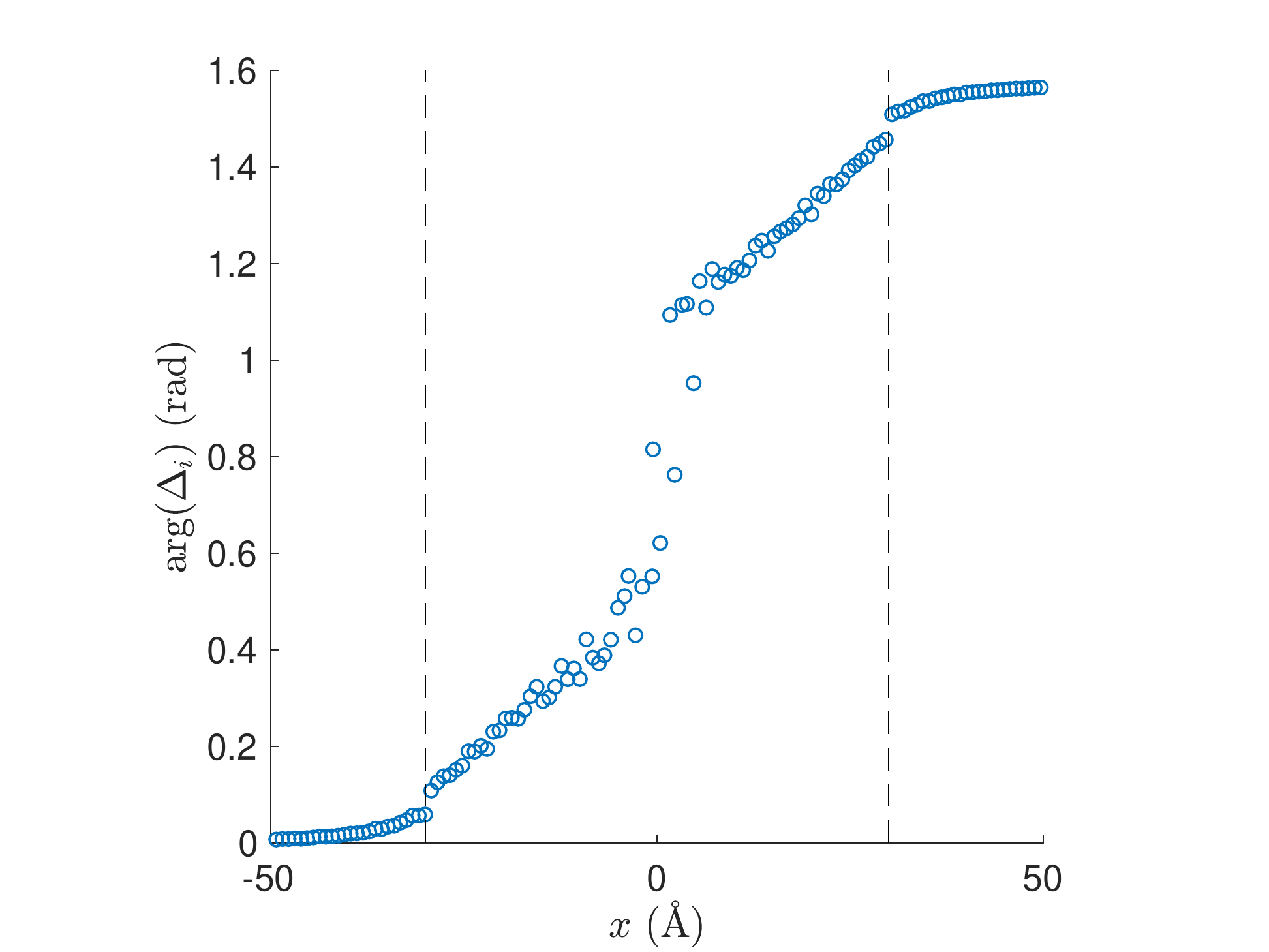}\\
	\caption{Plot of the order parameter phase vs. the distance along the $x$-axis for a symmetric (410) grain-boundary junction. The dashed lines indicate where the phase is held fixed ($x = \pm 30$ $\r{A}$). Note the sharp phase drop near the grain boundary, as well as the smooth gradient within the ordered grains. This phase drop implies that the current density through this system is smaller than that of a clean system without a grain boundary.}
	\label{fig:410gb_phase}
\end{figure}

\begin{figure}[ht]
	\centering
	\includegraphics[width=0.5\textwidth]{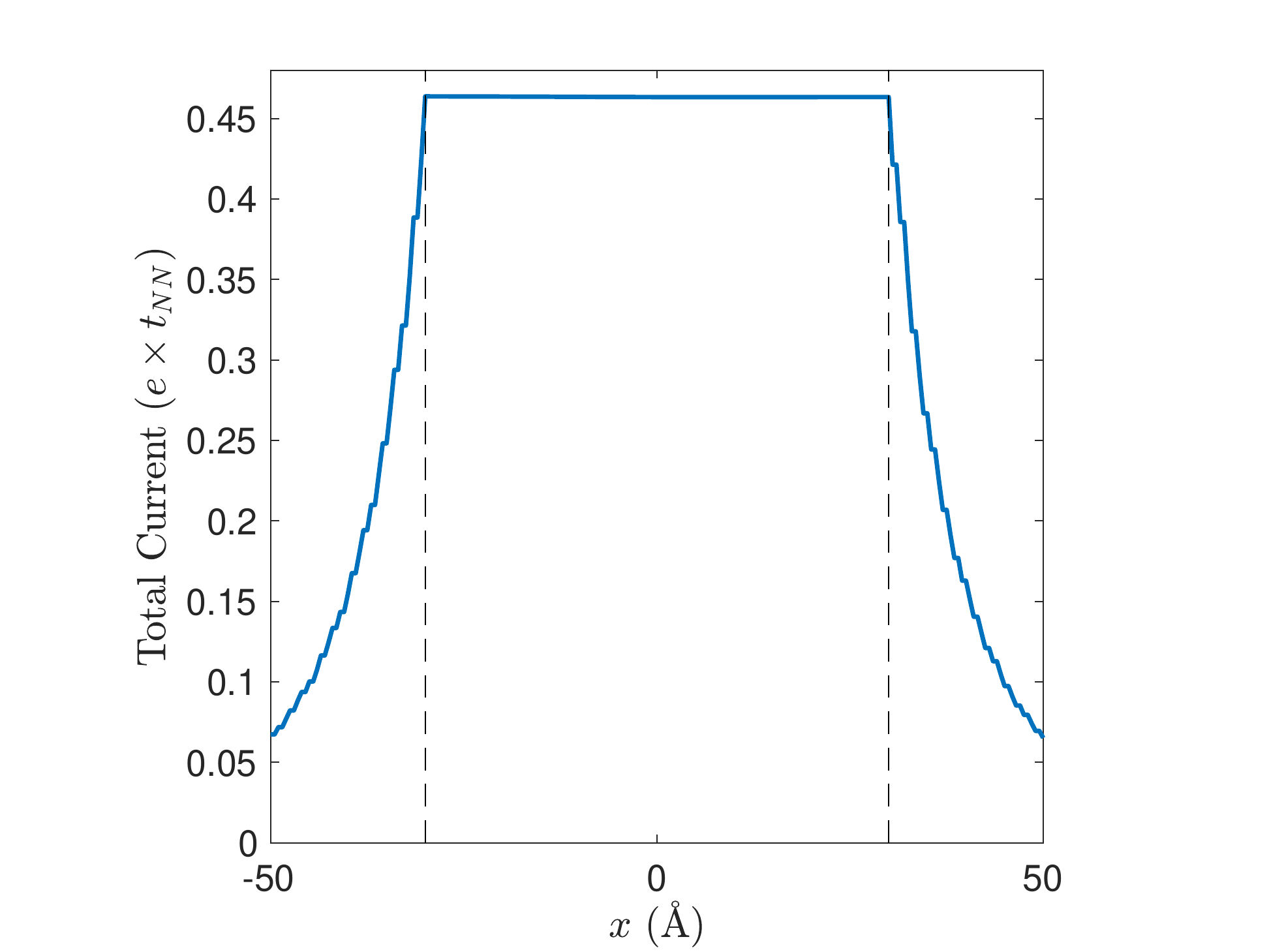}\\
	\caption{Plot of the total current (in units of $e \times t_\text{NN}$)  vs. the distance along the $x$-axis for a symmetric (410) grain-boundary junction. The dashed lines indicate where the phase is held fixed ($x = \pm 30$ $\r{A}$). Note that for $-30 \r{A} < x < 30 \r{A}$, the current is conserved (\emph{i.e.}, a constant with respect to $x$).}
	\label{fig:410gb_current}
\end{figure}

\begin{figure}[ht]
	\centering
	\includegraphics[width=0.5\textwidth]{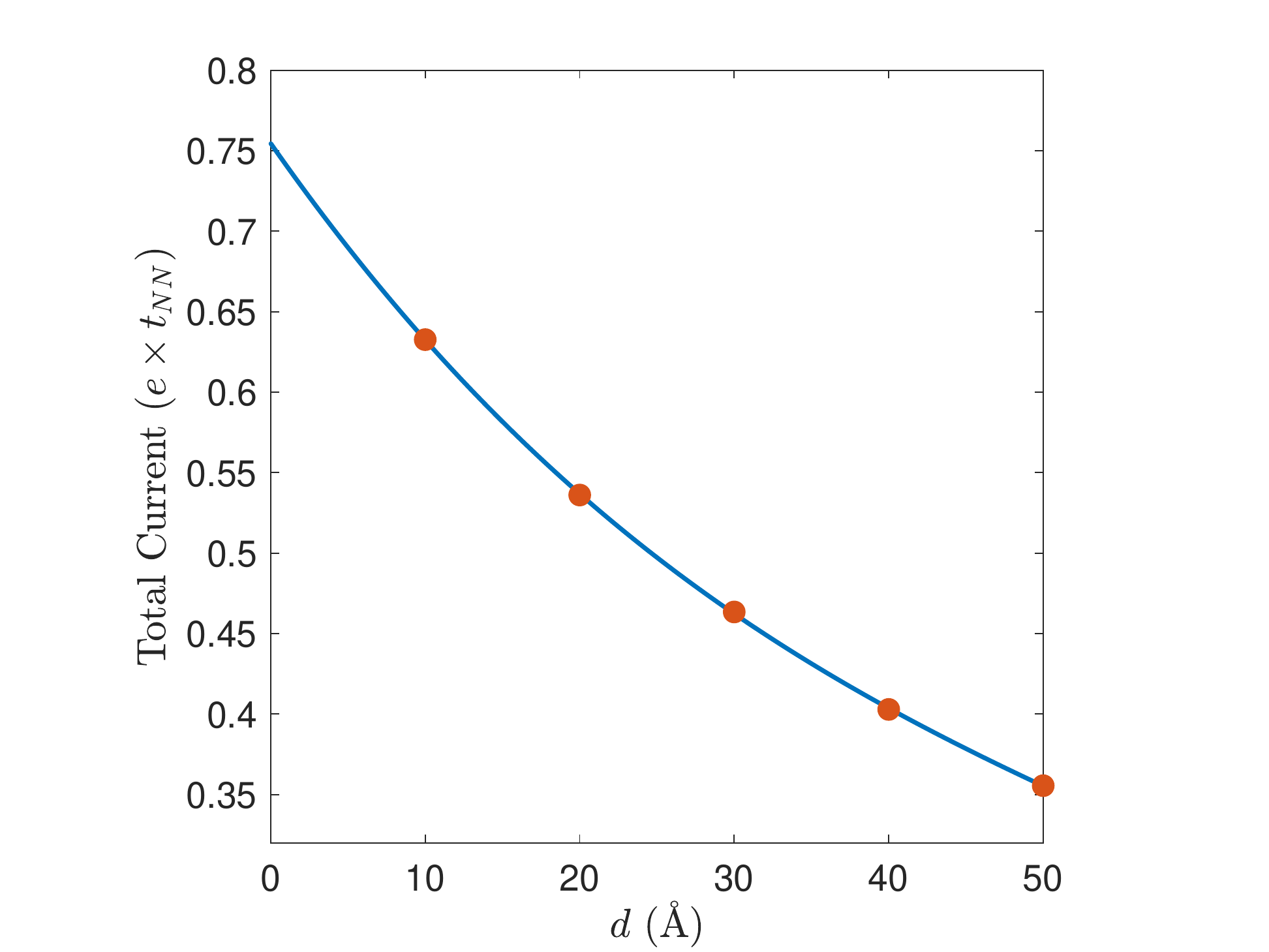}\\
	\caption{Plot of the total current (in units of $e \times t_\text{NN}$) vs. $d$ for a symmetric (410) grain-boundary junction together with extrapolation to $d=0$ to estimate the critical current. As $d$ is decreased, the phase gradient becomes larger, and consequently, the current increases. 
	}
	\label{fig:410gb_currentvsd}
\end{figure}

\begin{figure}[ht]
	\centering
	\includegraphics[width=0.9\linewidth]{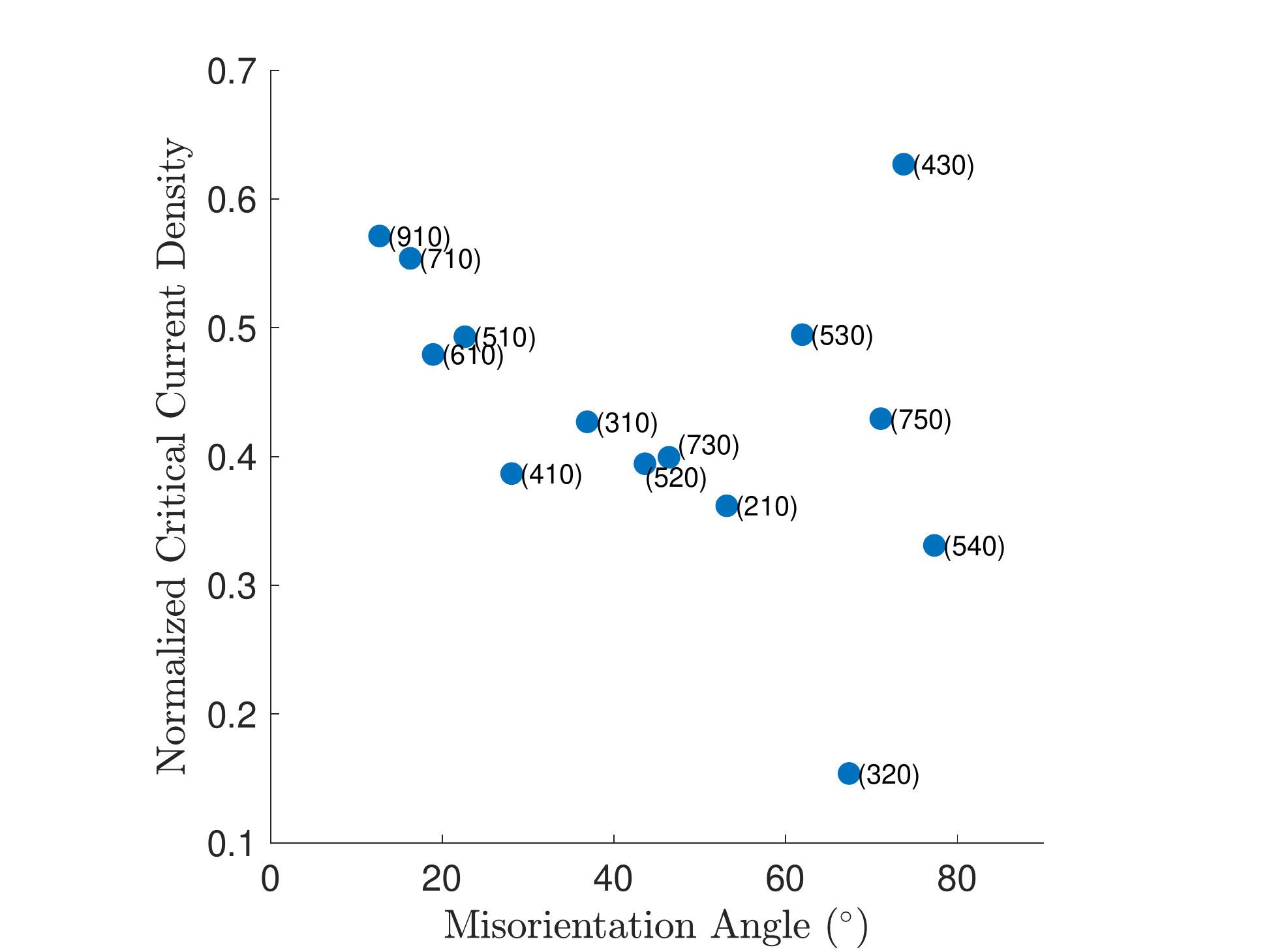}\\
	\caption{Plot of the critical-current density (normalized to that of the perfect-lattice case) vs. the misorientation angle for a variety of grain-boundary junctions. The critical-current density depends only weakly on the misorientation angle.} 
	\label{fig:currentvsangle}
\end{figure}

In this section, we will discuss the case where a grain boundary is placed in the middle of a superconducting system, and no tunnel barrier is present. Here, the grain boundary is oriented so that it is perpendicular to the phase gradient and hence the supercurrent flow.
This model may be of relevance to experimental studies of the critical current of polycrystalline Nb wires.\cite{rogachev2003superconducting}
In our simulations, we calculate the critical current for each of the two layers (along the $z$-direction) comprising the grain-boundary junction and take the average of the two critical currents; this effectively truncates the system down to a two-dimensional lattice, but because we have taken care to preserve the Fermi velocity and the Fermi-surface anisotropy of the full system within the single-band approximation we made for the electronic structure, many of the important qualitative features of the full three-dimensional system are retained even after performing this two-dimensional truncation. However, we take into account variations in the $z$-component of the atomic positions when computing the hopping amplitudes between sites (Fig. \ref{fig:410z}). In addition, following Graser \emph{et al.}, in calculating the critical current, we impose a phase gradient between $x =- d$ and $x = d$ and calculate the current for various values of $d$, and we take the extrapolated value of the current as $d \to 0$ to be the critical current. The reason for doing this is that in superconductor-only systems such as the grain-boundary junctions studied here, assuming that $d$ is fixed, the current flowing through the system will not be a periodic function of the phase difference between $x = -d$ and $x = d$ unlike the behavior expected for a junction with a normal component in the middle of the two leads. One thus needs to put various values of the phase gradient to sweep through the possible values of the current flowing through the system. In practice, we fix the phase difference between $x = -d$ and $x = d$ to be $\pi/2$ and vary $d$.\cite{graser2010grain} Strictly speaking, the $d \to 0$ limit is not achievable in practice as a fundamental limit is the minimum spacing between atoms, which is always nonzero, but one needs a measure of the critical current that is comparable across different grain-boundary types independent of the finer details of the system, and taking the $d \to 0$ limit provides one such way of achieving that. We use a cubic polynomial fit to obtain the value of the current in the $d \to 0$ limit. We choose a symmetric (410) grain boundary as a particular example to highlight the interesting physics present in such junctions, though none of these effects are specific to this particular junction, and all phenomena discussed here are found across all grain-boundary types. We adopt here the notation used in Ref.~\onlinecite{graser2010grain} to specify the grain boundary structure.

In Fig.~\ref{fig:410gb}, we show four  real-space plots of a symmetric (410) grain-boundary junction: the hopping amplitudes, the order parameter amplitude, the order parameter phase, and the bond current (\emph{i.e.}, the current between any two pairs of atoms). For this set of plots, the order parameter phase was held fixed at $x = -30$ \r{A} and $x = 30$ \r{A}.  It can be seen that the hopping amplitudes away from the grain boundary are ordered, with values consistent with what is expected from a perfect lattice. It is only when approaching the grain boundary that major changes are evident in the hoppings. There is a bond connecting two atoms from different grains on which the hopping amplitude is enhanced considerably relative to the ordered case, owing to the decreased distance between these two. There is another bond for which the hoppings are suppressed due to the increased interatomic distance between the two atoms. It is clear from  the map of hoppings that the grain boundary presents an obstruction to the clean flow of supercurrent between the left and right ends of the system; we will show later that much of the current within the grain boundary is in fact carried by a set of bonds for which the hoppings are neither too small nor too large.

The order parameter amplitude is sensitive to the disorder near the grain boundary, as can be seen in the real-space plot. Sites that are connected by enhanced hopping amplitudes show a suppressed order parameter amplitude. In addition, even the order parameter amplitudes on sites located slightly away from the grain boundary are affected by seemingly minute alterations to the atomic positions. Nevertheless, away from the grain boundary the order parameter acquires its clean-limit value, a sign that the effects of the grain boundary are highly localized only to nearby sites, as expected. The order parameter phase can be seen to be smoothly varying away from the grain boundary. However, near the grain boundary, a sharp drop can be seen in the phase moving from right to left: the phase varies very quickly across the grain boundary. This sharp phase drop is an effect of the hopping disorder within the grain boundary and can be visualized in clearer detail in Fig.~\ref{fig:410gb_phase}, which shows the phase as a function of the $x$-coordinate. The sharp phase drop means that the phase gradient within the ordered grains is smaller compared to the case where no grain boundary is present; hence the resulting current density will be smaller than in the clean limit.

The bond current changes from being ordered and flowing in a predictable left-to-right fashion within the ordered grains to being highly inhomogeneous near the grain boundary. In this region, much of the current is carried by the two 
bonds closest to $(x, y) = (0, 0)$ in Fig.~\ref{fig:410gb}d, 
 with comparatively little current flowing along bonds where the hoppings are suppressed. Interestingly, one can observe a number of reversals in the direction of the current, with some bonds showing current flowing from right to left (instead of the expected left to right). Because most of the current is carried by a small fraction of the bonds near the grain boundary, when current flows out from these bonds into the ordered portions of the junction, they tend to fill out and take up the available channels in the ordered grains, resulting in these local $\pi$-junctions where reversals of the current direction occur.

Despite the spatially inhomogeneous nature of the bond currents, current conservation still holds within the fully self-consistent evaluation, which is a nontrivial check on the validity of the calculation.\cite{furusaki1991dc,sols1994crossover} In Fig.~\ref{fig:410gb_current} we show the calculated \emph{total} current for a symmetric (410) grain-boundary junction as a function of the $x$-coordinate, which we obtain for each $x$ by summing together all bond currents between sites $i$ and $j$ such that $x_i < x \leq x_j$. Within the region where the phase gradient is well-defined (\emph{i.e.}, $-30$ \r{A} $< x < 30$ \r{A}), the total current is a constant function of $x$, an indication that current is conserved. In Fig.~\ref{fig:410gb_currentvsd}, we show how the total current for the (410) grain-boundary junction depends on $d$. As $d$ is decreased, the total current increases, which is expected since the current is determined by the phase gradient, which is inversely dependent on $d$ (recall that the phase difference is held fixed at $\pi/2$). We then extrapolate these to the $d \to 0$ limit to provide a measure of the critical current that is can be used to compare junctions with different grain-boundary types.

Because these systems are effectively two dimensional, we can define the \emph{critical-current density} as the total current in the $d \to 0$ limit divided by the width of the system. In Fig.~\ref{fig:currentvsangle}, we plot the critical-current density (normalized by the intrinsic critical-current density of a system with no grain boundary) versus the misorientation angle of the grain boundary. It can be seen that the critical-current density has only a weak dependence on the misorientation angle. The values of the critical-current density cluster within a narrow range, and regardless of the grain-boundary type, the suppression relative to the clean, perfect-lattice case is consistently large---typically around half of the clean limit. This weak angle-dependence can be contrasted with what is seen in the high-temperature cuprate superconductors, where an exponential dependence on the misorientation angle is observed instead.\cite{dimos1988orientation, hilgenkamp2002grain} As 
Nb is an $s$-wave superconductor and is not expected to exhibit local charge transfer effects that are crucial in inducing further suppression of the critical current in the cuprates\cite{graser2010grain}, it is not surprising that the angle-dependence of the critical current is much weaker for the cases studied here.

\section{Grain Boundaries In SIS Junctions}

\begin{figure*}[ht]
	\centering
	\includegraphics[width=\textwidth]{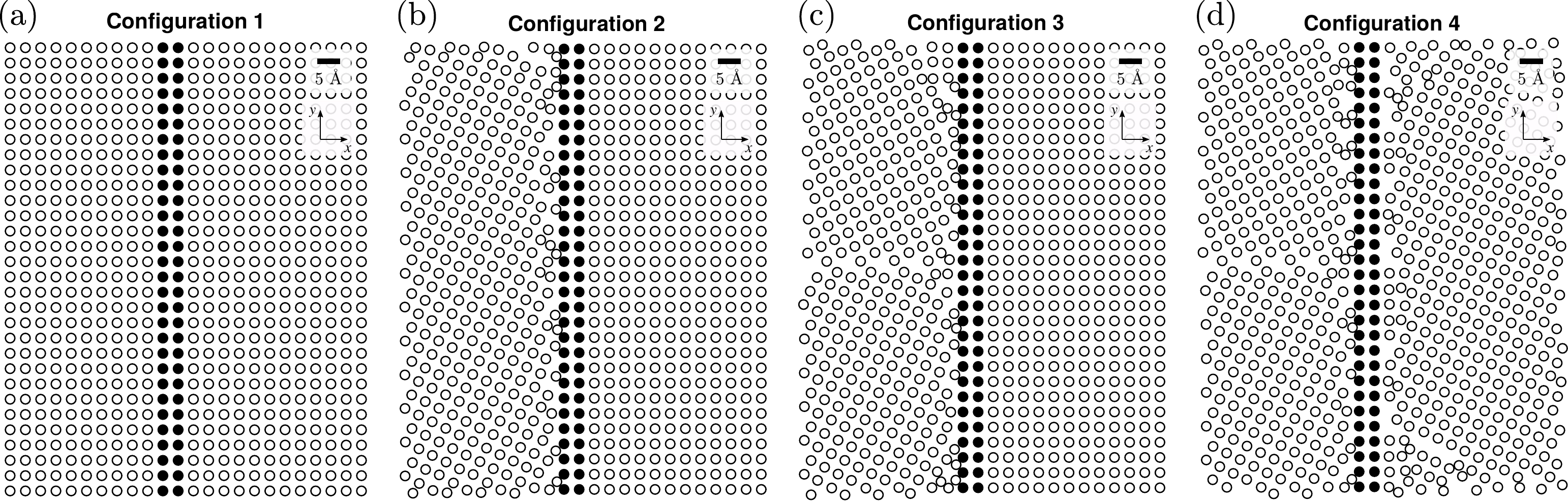}
	\caption{The four grain-boundary/tunnel barrier configurations studied (a-d). The black sites denote those which are part of the tunnel barrier; real-space scale and the coordinate axis direction are indicated in the insets.} 
	\label{fig:oxideconfigs}
\end{figure*}

\begin{figure}[ht]
	\centering
	\includegraphics[width=0.5\textwidth]{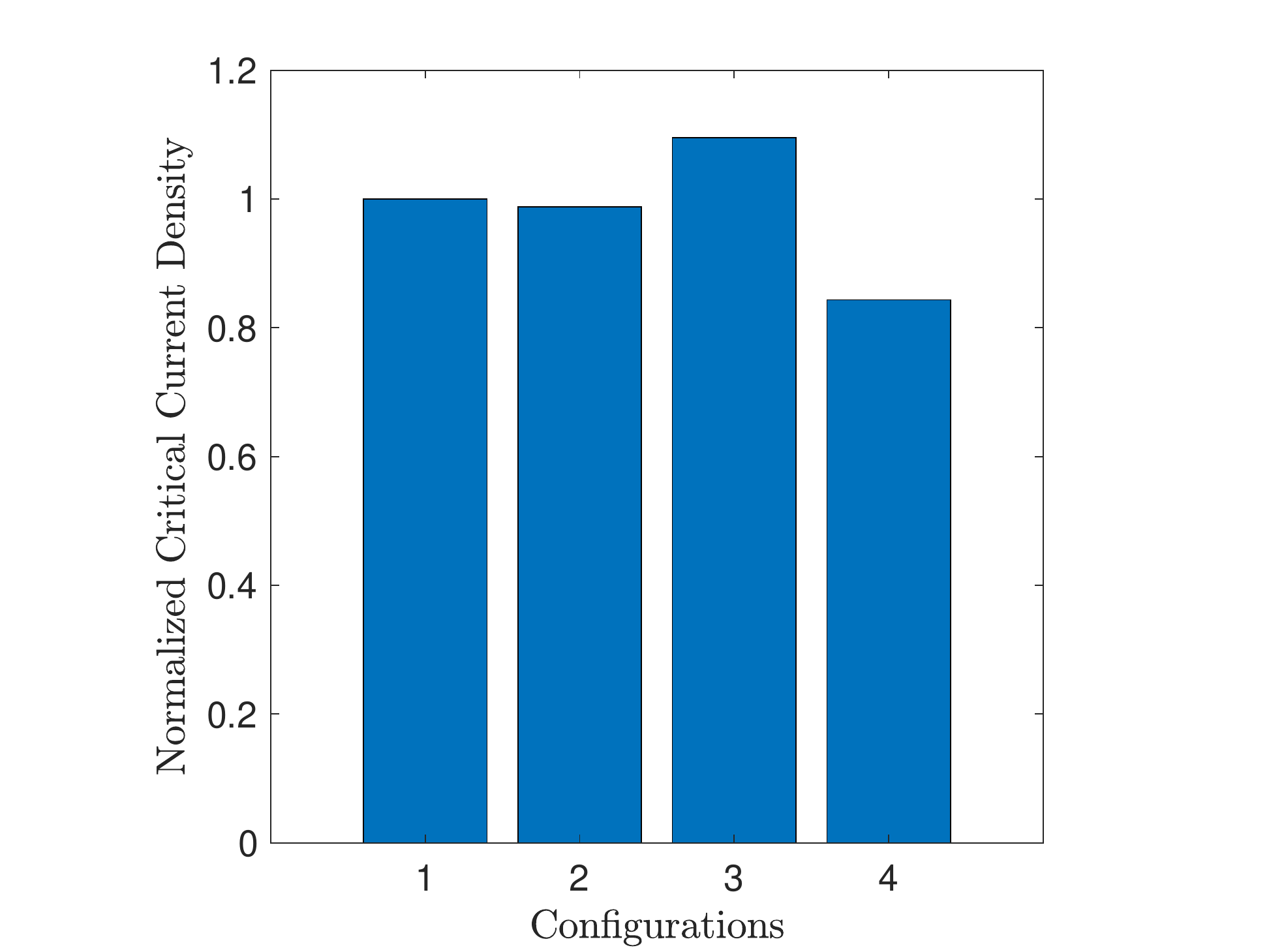}\\
	\caption{Plot of the normalized critical-current density for the various configurations shown in Fig.~\ref{fig:oxideconfigs}. The values are normalized to the critical-current density of Configuration 1.}
	\label{fig:ccconfigs}
\end{figure}

\begin{figure*}[ht]
	\centering
		\includegraphics[width=\textwidth]{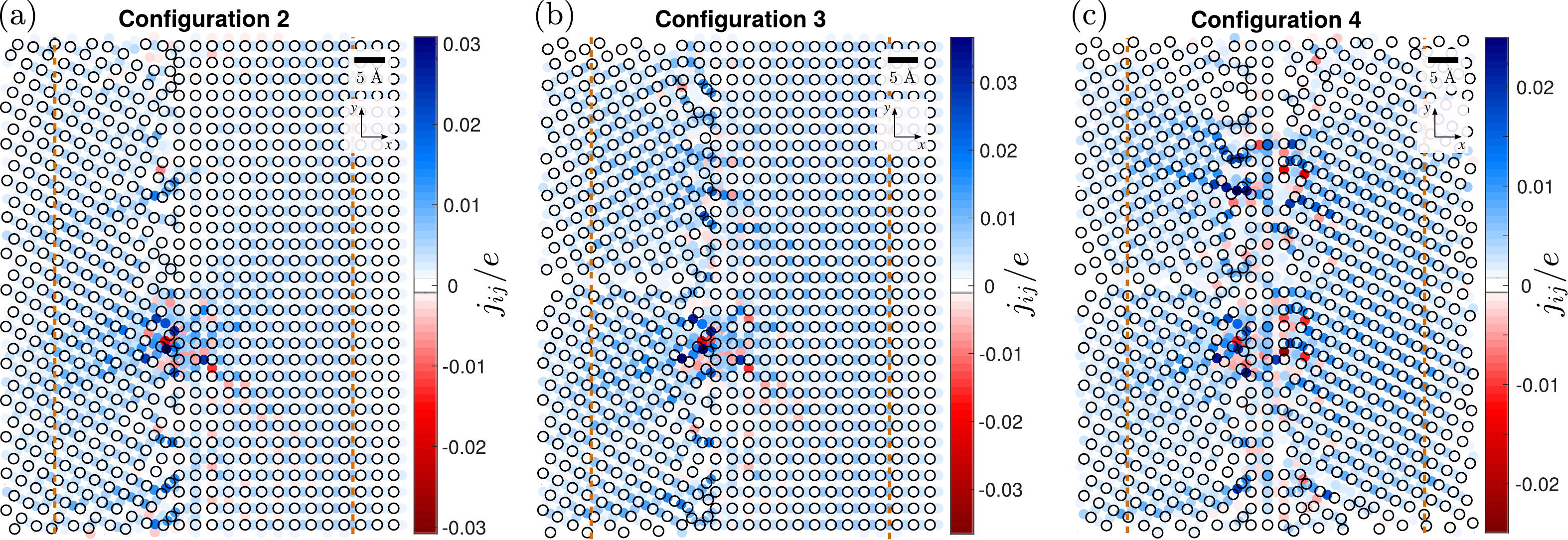}
	\caption{Plots of the bond currents for Configurations 2 (a), 3 (b), and 4 (c). The orange dashed lines indicate where the phase is held fixed ($x = \pm 30$ $\r{A}$). The real-space scalebar and the coordinate axis are indicated in the insets.}
	\label{fig:oxideconfigs_current}
\end{figure*}

\begin{figure}[ht]
	\centering
	\includegraphics[width=\linewidth]{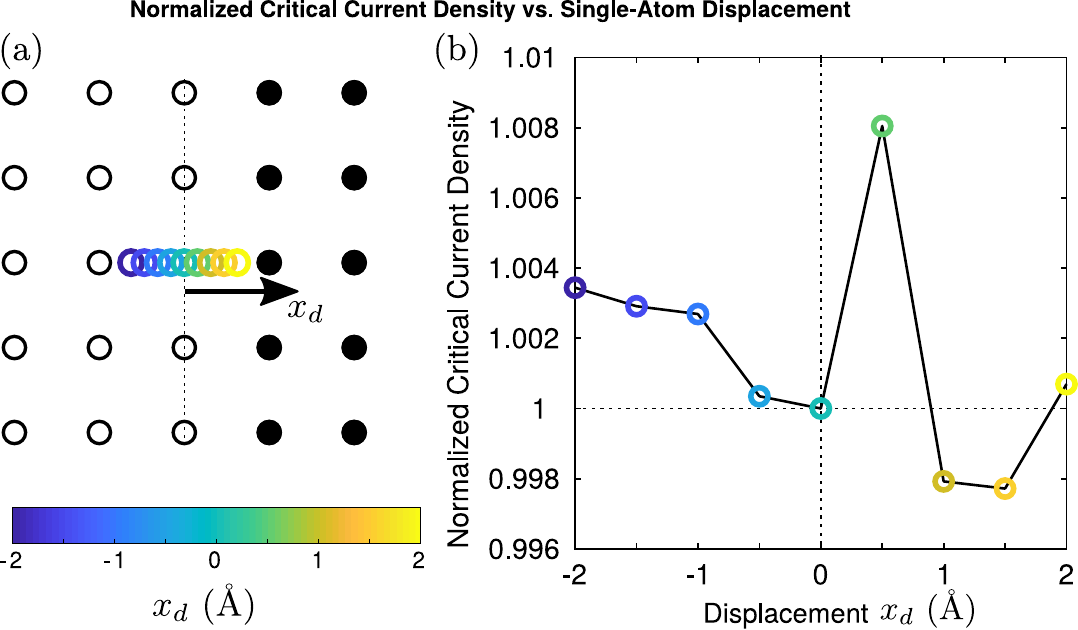}
	\caption{Plot of the normalized critical-current density vs. the single-atom displacement $x_d$ for Configuration 1. (a) The single-atom displacements $x_d$ used in the calculation, plotted within the surrounding lattice, and (b) the normalized critical-current density plotted as a function of $x_d$. The values are normalized to the critical-current density with displacement $x_d = 0$.}
	\label{fig:currentvsdisplacement}
\end{figure}

Having seen in the previous section that the disordered hoppings near a grain boundary naturally give rise to a large suppression of the critical current, 
we next discuss a similar, if slightly different context where grain boundaries also play a major role. Here, we consider a tunnel barrier placed between two superconducting leads so that the junction is of the superconductor-insulator-superconductor (SIS) type. We assume that the tunnel barrier is two layers thick. Given this basic SIS configuration, we can ask what happens if this SIS junction is disordered by the presence of grain boundaries. 

We  consider four different scenarios, as illustrated in Fig.~\ref{fig:oxideconfigs}.
The four GB boundary/tunnel barrier configurations shown in Fig.~\ref{fig:oxideconfigs} were set up and relaxed in a similar fashion employing LAMMPS. All sub-divisions of the simulation cell except the rigid tunnel barriers were subjected to rigid body translations along all GB planes present in the system, atom deletion criterion, and energy minimization using the same non-linear conjugate gradient energy method. To ensure periodic boundary conditions in the molecular dynamics simulations, we used two tunnel barriers. However, only one tunnel barrier is considered in the self-consistent BdG calculations, which employ open boundary conditions along the $x$-direction.

Configuration 1 is our baseline ordered case and consists of two superconducting leads which do not host grain boundaries within them and which are oriented identically to each other. Configuration 2 consists of left and right superconducting leads, which, while individually ordered in the sense of not hosting grain boundaries within each lead, are misoriented relative to each other. Configuration 3 consists of a left superconducting lead, which has a symmetric (210) grain boundary dividing its top and bottom halves and which is wholly misoriented relative to the right superconducting lead. The right lead is otherwise ordered and has no grain boundary. Last, Configuration 4 is the same as Configuration 3 but has the right superconducting lead oriented identically to the bottom half of the left superconducting lead. The tunnel barrier layers are assumed to be ordered and fixed. To calculate the critical current, we set the phase difference to $\pi/2$ (where a maximum in the current occurs for a generic SIS junction) and hold the phase fixed at $x = -30$ \r{A} and $x = 30$ \r{A}. There is no need to perform a sweep through $d$ since a normal component (the insulating barrier) is present, and for junctions of this type the phase gradient is a meaningful quantity only throughout the normal portion of the system; \emph{i.e.}, the critical-current density is independent of the choice of $d$. One set of $x = \pm d$ values is as good as any other, as long as these encompass the normal portion in its entirety.

It is worth noting why we choose to focus on  these four particular configurations and not on a larger ensemble of configurations. At present,  we do not have any concrete information as to the presence, location, and distribution of GBs near the junction in individual Nb junctions or on the variance of these factors over the thousands of junctions fabricated in modern processes.  Transmission electron microscopy (TEM) imaging, which would assist in these determinations, is not available.  Unlike thickness variations, the density of grain boundaries alone cannot be estimated from easily repeated measurements across an ensemble of junctions.  In the absence of such guides, we have adopted the approach of considering representative types of GBs that might occur near a junction, without considering orientation, to obtain a qualitative sense of how the presence of a GB affects the critical current relative to the case with a simple junction but no GBs.  Calculating large numbers of grain boundary configurations thus appears to us not productive at present. We anticipate that our calculations will be useful to analyze critical current variability when detailed TEM images become available.

Our main results are shown in Fig.~\ref{fig:ccconfigs}, which shows the critical-current density for all four configurations, normalized by that of Configuration 1 (the ordered case). It is evident that variations are present in the critical-current density, which do not necessarily reflect a monotonic increase in the disorder level that might naively be expected in moving from Configuration 1 to 4. Configurations 1 and 2 have nearly the same critical-current density, but Configuration 3 has a \emph{higher} critical-current density than either (by around 10\%), and Configuration 4 has a \emph{lower} one than the baseline level (by around 15\%). 

This variability is not surprising, but it is worth noting that much of it is due to the highly disordered nature of the hoppings near the tunnel barrier for Configurations 2-4. When hopping disorder is present, the transport through the tunnel barrier  is no longer homogeneous, as in Configuration 1. Rather, the current follows paths through the system where the hopping amplitudes near the tunnel barrier are large, and for Configurations 2-4, there exist portions of the cross section near the tunnel barrier which do not host any supercurrent at all. Much of this depends on a fortuitous combination of pathways that allow current to flow as unimpeded as possible. We illustrate this in Fig.~\ref{fig:oxideconfigs_current}, which shows the bond currents for Configurations 2, 3, and 4. It can be seen that the reason Configuration 2 has a smaller critical current than Configuration 3, even though the bottom halves of both systems are the same, is because for Configuration 2, the top half of the left superconducting lead has only a very small number of viable transport channels for current to flow, while Configuration 3's top half of the left lead features a set of bonds near the tunnel barrier whose hopping amplitudes prove to be optimal for transport owing to their large amplitudes. Configuration 4 in turn has a smaller critical current than Configuration 3, despite having identical left superconducting leads, primarily because Configuration 4's \emph{right} superconducting lead has a much more disordered set of hoppings near the tunnel barrier, providing an additional obstruction to the flow of the current that was otherwise not present in Configuration 3. Evidently, from these examples, the role of a fortunate set of atomic locations in ensuring optimal current flow cannot be understated.

As an additional demonstration of the importance of even small shifts in atomic positions on the critical-current density, we perform a calculation involving Configuration 1, but with \emph{one} atom on the left-hand side of the tunnel barrier displaced from its ordered position along the $x$-axis. We calculate the critical current for different atomic displacements; the results are shown in Fig.~\ref{fig:currentvsdisplacement}, which shows the critical-current density normalized relative to the case where \emph{no} displacement is present versus the displacement $x_d$. We can see first that the differences in the critical current are very small, which is not surprising given that only one atom is being shifted in these calculations. For much of the displacements, the change in the critical current is small---around 0.03\% to 0.2\%. However, for $x_d = 0.5$ \r{A}, the enhancement turns out to be quite large at around 0.8\%. It turns out that at this value of the displacement, the hoppings are close to the optimal value to facilitate current flow. If the displacement were to be made larger such that the atom becomes very close to the atom to its left or right, the hoppings can actually become too large on one side and too small on the other, and inhibit current flow through those bonds such that the displaced atom does not participate in transport any more, and the current pattern merely shifts to go around the non-participating atom. However, for the $x_d = 0.5$ \r{A} case, the displacement is optimized enough for the atom to collect current from the atoms immediately above and below it without having the hopping amplitude be suppressed enough for that particular transport process to occur. (For $x_d = 0$ \r{A}, that transport process does not occur since the non-shifted atom and the atoms above and below it are at the same phase, and therefore, there is no current flow.) This additional transport process, made possible by the right combination of locations and the resulting hopping amplitudes, turns out to contribute a small but non-negligible enhancement to the perfectly ordered case, which is much larger than the variations one can obtain from other values of atomic displacement (such as $x_d = 1.0$ \r{A} or $x_d = -0.5$ \r{A}), for which the resulting critical-current shift is much smaller. 

We note too that even though the effect is small on the level of one atom being displaced, the effect can be pronounced when a large number of atoms are displaced, as evidenced by our earlier discussion on Configurations 2, 3, and 4. Our results thus suggest that the hopping disorder that is naturally generated by grain boundaries can indeed give rise to the variations seen in experiment. Grain boundaries thus are an additional source of disorder that can lead to the observed device-to-device variation in the critical current in addition to the oxide thickness variations and vacancies that we previously studied as possible sources of this variation.

\section{Discussion and Conclusion}

In this paper, we have discussed the impact of grain boundaries on the critical current of Josephson junctions. We have seen that in a junction consisting of a grain boundary oriented perpendicularly to the phase gradient, there is a reduction in the critical current relative to the clean case, arising from the disordered hoppings in the vicinity of the grain boundary which cause a sharp phase drop within that region. These effects are largely independent of the particulars of the grain boundary. The minimal ingredients for this critical-current suppression are merely the suppression of hopping amplitudes near the grain boundary, which are enough to pose an obstacle to the smooth flow of supercurrent that would otherwise be the case without a grain boundary. We have also found that in the more physically relevant case where grain boundaries disorder an SIS tunnel junction, the highly disordered hopping amplitudes between the superconducting leads and the tunnel barrier lead to critical-current variations that are extremely sensitive to the precise position of the atoms. We find from an analysis of a small number of configurations that these variations can potentially be large. Our work demonstrates a physically plausible mechanism for critical-current variability that is distinct from previously considered explanations such as thickness variations.

Our work demonstrates that the presence of even ostensibly mild forms of disorder, such as small shifts in the atomic positions, can have a prominent effect on an observable such as the critical current. This marked sensitivity to atomic positions makes it all the more important to understand just how much disorder due to grain boundaries is present in real-world Nb/Al-AlO$_x$/Nb junctions. Much of this information is not easily accessible, owing perhaps to the  difficulty of performing transmission electron microscopy experiments on thousands of Josephson junctions, but is important if one wants to isolate the possible types of disorder responsible for the variability. In the absence of such images, our work still provides a valuable clue: the variability of the critical current due to grain boundaries is potentially large,
suggesting that efforts to control their occurrence in the fabrication
process are crucial.

\begin{acknowledgments}
	 P.J.H. is grateful for useful conversations with R. Hennig.
M.A.S., T.A.W., N.P., E.P., and M.L. acknowledge support from the IARPA SuperTools program. M.A.S. and P.J.H. acknowledge support from NSF-DMR-1849751. 
 
\end{acknowledgments}

\section*{Author Declarations}

\subsection{Conflict of Interest}

The authors have no conflicts to disclose.

\subsection{Ethics Approval}

No experiments on animal or human subjects were used for the preparation of the submitted manuscript.

\section*{Data Availability Statement}

The data that support the findings of this study are available from the corresponding author upon reasonable request.


\bibliography{paper_jj_gb}

\providecommand{\noopsort}[1]{}\providecommand{\singleletter}[1]{#1}%
\begin{thebibliography}{43}%
\makeatletter
\providecommand \@ifxundefined [1]{%
 \@ifx{#1\undefined}
}%
\providecommand \@ifnum [1]{%
 \ifnum #1\expandafter \@firstoftwo
 \else \expandafter \@secondoftwo
 \fi
}%
\providecommand \@ifx [1]{%
 \ifx #1\expandafter \@firstoftwo
 \else \expandafter \@secondoftwo
 \fi
}%
\providecommand \natexlab [1]{#1}%
\providecommand \enquote  [1]{``#1''}%
\providecommand \bibnamefont  [1]{#1}%
\providecommand \bibfnamefont [1]{#1}%
\providecommand \citenamefont [1]{#1}%
\providecommand \href@noop [0]{\@secondoftwo}%
\providecommand \href [0]{\begingroup \@sanitize@url \@href}%
\providecommand \@href[1]{\@@startlink{#1}\@@href}%
\providecommand \@@href[1]{\endgroup#1\@@endlink}%
\providecommand \@sanitize@url [0]{\catcode `\\12\catcode `\$12\catcode
  `\&12\catcode `\#12\catcode `\^12\catcode `\_12\catcode `\%12\relax}%
\providecommand \@@startlink[1]{}%
\providecommand \@@endlink[0]{}%
\providecommand \url  [0]{\begingroup\@sanitize@url \@url }%
\providecommand \@url [1]{\endgroup\@href {#1}{\urlprefix }}%
\providecommand \urlprefix  [0]{URL }%
\providecommand \Eprint [0]{\href }%
\providecommand \doibase [0]{http://dx.doi.org/}%
\providecommand \selectlanguage [0]{\@gobble}%
\providecommand \bibinfo  [0]{\@secondoftwo}%
\providecommand \bibfield  [0]{\@secondoftwo}%
\providecommand \translation [1]{[#1]}%
\providecommand \BibitemOpen [0]{}%
\providecommand \bibitemStop [0]{}%
\providecommand \bibitemNoStop [0]{.\EOS\space}%
\providecommand \EOS [0]{\spacefactor3000\relax}%
\providecommand \BibitemShut  [1]{\csname bibitem#1\endcsname}%
\let\auto@bib@innerbib\@empty
\bibitem [{\citenamefont {{McCaughan}}\ and\ \citenamefont
  {{Berggren}}(2014)}]{McCaughan2014}%
  \BibitemOpen
  \bibfield  {author} {\bibinfo {author} {\bibfnamefont {A.~N.}\ \bibnamefont
  {{McCaughan}}}\ and\ \bibinfo {author} {\bibfnamefont {K.~K.}\ \bibnamefont
  {{Berggren}}},\ }\href {\doibase 10.1021/nl502629x} {\bibfield  {journal}
  {\bibinfo  {journal} {Nano Letters}\ }\textbf {\bibinfo {volume} {14}},\
  \bibinfo {pages} {5748} (\bibinfo {year} {2014})},\ \Eprint
  {http://arxiv.org/abs/1403.6423} {arXiv:1403.6423 [cond-mat.supr-con]}
  \BibitemShut {NoStop}%
\bibitem [{\citenamefont {Tolpygo}(2016)}]{tolpygo2016superconductor}%
  \BibitemOpen
  \bibfield  {author} {\bibinfo {author} {\bibfnamefont {S.~K.}\ \bibnamefont
  {Tolpygo}},\ }\href@noop {} {\bibfield  {journal} {\bibinfo  {journal} {Low
  Temperature Physics}\ }\textbf {\bibinfo {volume} {42}},\ \bibinfo {pages}
  {361} (\bibinfo {year} {2016})}\BibitemShut {NoStop}%
\bibitem [{\citenamefont {Gurvitch}\ \emph {et~al.}(1983)\citenamefont
  {Gurvitch}, \citenamefont {Washington},\ and\ \citenamefont
  {Huggins}}]{gurvitch1983high}%
  \BibitemOpen
  \bibfield  {author} {\bibinfo {author} {\bibfnamefont {M.}~\bibnamefont
  {Gurvitch}}, \bibinfo {author} {\bibfnamefont {M.~A.}\ \bibnamefont
  {Washington}}, \ and\ \bibinfo {author} {\bibfnamefont {H.~A.}\ \bibnamefont
  {Huggins}},\ }\href@noop {} {\bibfield  {journal} {\bibinfo  {journal}
  {Applied Physics Letters}\ }\textbf {\bibinfo {volume} {42}},\ \bibinfo
  {pages} {472} (\bibinfo {year} {1983})}\BibitemShut {NoStop}%
\bibitem [{\citenamefont {Tolpygo}\ \emph {et~al.}(2015)\citenamefont
  {Tolpygo}, \citenamefont {Bolkhovsky}, \citenamefont {Weir}, \citenamefont
  {Johnson}, \citenamefont {Gouker},\ and\ \citenamefont
  {Oliver}}]{tolpygo2014fabrication}%
  \BibitemOpen
  \bibfield  {author} {\bibinfo {author} {\bibfnamefont {S.~K.}\ \bibnamefont
  {Tolpygo}}, \bibinfo {author} {\bibfnamefont {V.}~\bibnamefont {Bolkhovsky}},
  \bibinfo {author} {\bibfnamefont {T.~J.}\ \bibnamefont {Weir}}, \bibinfo
  {author} {\bibfnamefont {L.~M.}\ \bibnamefont {Johnson}}, \bibinfo {author}
  {\bibfnamefont {M.~A.}\ \bibnamefont {Gouker}}, \ and\ \bibinfo {author}
  {\bibfnamefont {W.~D.}\ \bibnamefont {Oliver}},\ }\href@noop {} {\bibfield
  {journal} {\bibinfo  {journal} {IEEE Transactions on Applied
  Superconductivity}\ }\textbf {\bibinfo {volume} {25}},\ \bibinfo {pages} {1}
  (\bibinfo {year} {2015})},\ \bibinfo {note} {1101312},\ \Eprint
  {http://arxiv.org/abs/https://ieeexplore.ieee.org/document/6967749}
  {https://ieeexplore.ieee.org/document/6967749} \BibitemShut {NoStop}%
\bibitem [{\citenamefont {Holmes}\ and\ \citenamefont
  {McHenry}(2017)}]{holmes2017non}%
  \BibitemOpen
  \bibfield  {author} {\bibinfo {author} {\bibfnamefont {D.~S.}\ \bibnamefont
  {Holmes}}\ and\ \bibinfo {author} {\bibfnamefont {J.}~\bibnamefont
  {McHenry}},\ }\href@noop {} {\bibfield  {journal} {\bibinfo  {journal} {IEEE
  Transactions on Applied Superconductivity}\ }\textbf {\bibinfo {volume}
  {27}},\ \bibinfo {pages} {1} (\bibinfo {year} {2017})},\ \bibinfo {note}
  {1100605},\ \Eprint
  {http://arxiv.org/abs/https://ieeexplore.ieee.org/document/7792220}
  {https://ieeexplore.ieee.org/document/7792220} \BibitemShut {NoStop}%
\bibitem [{\citenamefont {Tolpygo}\ \emph {et~al.}(2017)\citenamefont
  {Tolpygo}, \citenamefont {Bolkhovsky}, \citenamefont {Zarr}, \citenamefont
  {Weir}, \citenamefont {Wynn}, \citenamefont {Day}, \citenamefont {Johnson},\
  and\ \citenamefont {Gouker}}]{tolpygo2017properties}%
  \BibitemOpen
  \bibfield  {author} {\bibinfo {author} {\bibfnamefont {S.~K.}\ \bibnamefont
  {Tolpygo}}, \bibinfo {author} {\bibfnamefont {V.}~\bibnamefont {Bolkhovsky}},
  \bibinfo {author} {\bibfnamefont {S.}~\bibnamefont {Zarr}}, \bibinfo {author}
  {\bibfnamefont {T.}~\bibnamefont {Weir}}, \bibinfo {author} {\bibfnamefont
  {A.}~\bibnamefont {Wynn}}, \bibinfo {author} {\bibfnamefont {A.~L.}\
  \bibnamefont {Day}}, \bibinfo {author} {\bibfnamefont {L.}~\bibnamefont
  {Johnson}}, \ and\ \bibinfo {author} {\bibfnamefont {M.}~\bibnamefont
  {Gouker}},\ }\href@noop {} {\bibfield  {journal} {\bibinfo  {journal} {IEEE
  Transactions on Applied Superconductivity}\ }\textbf {\bibinfo {volume}
  {27}},\ \bibinfo {pages} {1} (\bibinfo {year} {2017})},\ \bibinfo {note}
  {1100815},\ \Eprint
  {http://arxiv.org/abs/https://ieeexplore.ieee.org/document/7851016}
  {https://ieeexplore.ieee.org/document/7851016} \BibitemShut {NoStop}%
\bibitem [{\citenamefont {Tolpygo}\ \emph {et~al.}(2019)\citenamefont
  {Tolpygo}, \citenamefont {Bolkhovsky}, \citenamefont {Rastogi}, \citenamefont
  {Zarr}, \citenamefont {Day}, \citenamefont {Golden}, \citenamefont {Weir},
  \citenamefont {Wynn},\ and\ \citenamefont {Johnson}}]{tolpygo2019advanced}%
  \BibitemOpen
  \bibfield  {author} {\bibinfo {author} {\bibfnamefont {S.~K.}\ \bibnamefont
  {Tolpygo}}, \bibinfo {author} {\bibfnamefont {V.}~\bibnamefont {Bolkhovsky}},
  \bibinfo {author} {\bibfnamefont {R.}~\bibnamefont {Rastogi}}, \bibinfo
  {author} {\bibfnamefont {S.}~\bibnamefont {Zarr}}, \bibinfo {author}
  {\bibfnamefont {A.~L.}\ \bibnamefont {Day}}, \bibinfo {author} {\bibfnamefont
  {E.}~\bibnamefont {Golden}}, \bibinfo {author} {\bibfnamefont {T.~J.}\
  \bibnamefont {Weir}}, \bibinfo {author} {\bibfnamefont {A.}~\bibnamefont
  {Wynn}}, \ and\ \bibinfo {author} {\bibfnamefont {L.~M.}\ \bibnamefont
  {Johnson}},\ }\href@noop {} {\bibfield  {journal} {\bibinfo  {journal} {IEEE
  Transactions on Applied Superconductivity}\ }\textbf {\bibinfo {volume}
  {29}},\ \bibinfo {pages} {1} (\bibinfo {year} {2019})},\ \bibinfo {note}
  {1102513},\ \Eprint
  {http://arxiv.org/abs/https://ieeexplore.ieee.org/document/8666758}
  {https://ieeexplore.ieee.org/document/8666758} \BibitemShut {NoStop}%
\bibitem [{\citenamefont {Sulangi}\ \emph {et~al.}(2020)\citenamefont
  {Sulangi}, \citenamefont {Weingartner}, \citenamefont {Pokhrel},
  \citenamefont {Patrick}, \citenamefont {Law},\ and\ \citenamefont
  {Hirschfeld}}]{sulangi2020disorder}%
  \BibitemOpen
  \bibfield  {author} {\bibinfo {author} {\bibfnamefont {M.~A.}\ \bibnamefont
  {Sulangi}}, \bibinfo {author} {\bibfnamefont {T.~A.}\ \bibnamefont
  {Weingartner}}, \bibinfo {author} {\bibfnamefont {N.}~\bibnamefont
  {Pokhrel}}, \bibinfo {author} {\bibfnamefont {E.}~\bibnamefont {Patrick}},
  \bibinfo {author} {\bibfnamefont {M.}~\bibnamefont {Law}}, \ and\ \bibinfo
  {author} {\bibfnamefont {P.~J.}\ \bibnamefont {Hirschfeld}},\ }\href@noop {}
  {\bibfield  {journal} {\bibinfo  {journal} {Journal of Applied Physics}\
  }\textbf {\bibinfo {volume} {127}},\ \bibinfo {pages} {033901} (\bibinfo
  {year} {2020})}\BibitemShut {NoStop}%
\bibitem [{\citenamefont {Pokhrel}\ \emph {et~al.}(2021)\citenamefont
  {Pokhrel}, \citenamefont {Weingartner}, \citenamefont {Sulangi},
  \citenamefont {Patrick},\ and\ \citenamefont {Law}}]{pokhrel2021modeling}%
  \BibitemOpen
  \bibfield  {author} {\bibinfo {author} {\bibfnamefont {N.}~\bibnamefont
  {Pokhrel}}, \bibinfo {author} {\bibfnamefont {T.~A.}\ \bibnamefont
  {Weingartner}}, \bibinfo {author} {\bibfnamefont {M.~A.}\ \bibnamefont
  {Sulangi}}, \bibinfo {author} {\bibfnamefont {E.~E.}\ \bibnamefont
  {Patrick}}, \ and\ \bibinfo {author} {\bibfnamefont {M.~E.}\ \bibnamefont
  {Law}},\ }\href@noop {} {\bibfield  {journal} {\bibinfo  {journal} {IEEE
  Transactions on Applied Superconductivity}\ }\textbf {\bibinfo {volume}
  {31}},\ \bibinfo {pages} {1} (\bibinfo {year} {2021})},\ \bibinfo {note}
  {1101305},\ \Eprint
  {http://arxiv.org/abs/https://ieeexplore.ieee.org/document/9380150}
  {https://ieeexplore.ieee.org/document/9380150} \BibitemShut {NoStop}%
\bibitem [{\citenamefont {Santhanam}(1976)}]{santhanam1976flux}%
  \BibitemOpen
  \bibfield  {author} {\bibinfo {author} {\bibfnamefont {A.}~\bibnamefont
  {Santhanam}},\ }\href@noop {} {\bibfield  {journal} {\bibinfo  {journal}
  {Journal of Materials Science}\ }\textbf {\bibinfo {volume} {11}},\ \bibinfo
  {pages} {1099} (\bibinfo {year} {1976})}\BibitemShut {NoStop}%
\bibitem [{\citenamefont {DasGupta}\ \emph {et~al.}(1978)\citenamefont
  {DasGupta}, \citenamefont {Koch}, \citenamefont {Kroeger},\ and\
  \citenamefont {Chou}}]{dasgupta1978flux}%
  \BibitemOpen
  \bibfield  {author} {\bibinfo {author} {\bibfnamefont {A.}~\bibnamefont
  {DasGupta}}, \bibinfo {author} {\bibfnamefont {C.~C.}\ \bibnamefont {Koch}},
  \bibinfo {author} {\bibfnamefont {D.~M.}\ \bibnamefont {Kroeger}}, \ and\
  \bibinfo {author} {\bibfnamefont {Y.~T.}\ \bibnamefont {Chou}},\ }\href@noop
  {} {\bibfield  {journal} {\bibinfo  {journal} {Philosophical Magazine B}\
  }\textbf {\bibinfo {volume} {38}},\ \bibinfo {pages} {367} (\bibinfo {year}
  {1978})}\BibitemShut {NoStop}%
\bibitem [{\citenamefont {Zerweck}(1981)}]{zerweck1981pinning}%
  \BibitemOpen
  \bibfield  {author} {\bibinfo {author} {\bibfnamefont {G.}~\bibnamefont
  {Zerweck}},\ }\href@noop {} {\bibfield  {journal} {\bibinfo  {journal}
  {Journal of Low Temperature Physics}\ }\textbf {\bibinfo {volume} {42}},\
  \bibinfo {pages} {1} (\bibinfo {year} {1981})},\ \Eprint
  {http://arxiv.org/abs/https://doi.org/10.1007/BF00116692}
  {https://doi.org/10.1007/BF00116692} \BibitemShut {NoStop}%
\bibitem [{\citenamefont {Garg}\ \emph
  {et~al.}(2020{\natexlab{a}})\citenamefont {Garg}, \citenamefont {Muhich},
  \citenamefont {Cooley}, \citenamefont {Bieler},\ and\ \citenamefont
  {Solanki}}]{garg2020possible}%
  \BibitemOpen
  \bibfield  {author} {\bibinfo {author} {\bibfnamefont {P.}~\bibnamefont
  {Garg}}, \bibinfo {author} {\bibfnamefont {C.}~\bibnamefont {Muhich}},
  \bibinfo {author} {\bibfnamefont {L.~D.}\ \bibnamefont {Cooley}}, \bibinfo
  {author} {\bibfnamefont {T.~R.}\ \bibnamefont {Bieler}}, \ and\ \bibinfo
  {author} {\bibfnamefont {K.~N.}\ \bibnamefont {Solanki}},\ }\href@noop {}
  {\bibfield  {journal} {\bibinfo  {journal} {Phys. Rev. B}\ }\textbf {\bibinfo
  {volume} {101}},\ \bibinfo {pages} {184102} (\bibinfo {year}
  {2020}{\natexlab{a}})}\BibitemShut {NoStop}%
\bibitem [{\citenamefont {Nik}\ \emph {et~al.}(2016)\citenamefont {Nik},
  \citenamefont {Krantz}, \citenamefont {Zeng}, \citenamefont {Greibe},
  \citenamefont {Pettersson}, \citenamefont {Gustafsson}, \citenamefont
  {Delsing},\ and\ \citenamefont {Olsson}}]{nik2016correlation}%
  \BibitemOpen
  \bibfield  {author} {\bibinfo {author} {\bibfnamefont {S.}~\bibnamefont
  {Nik}}, \bibinfo {author} {\bibfnamefont {P.}~\bibnamefont {Krantz}},
  \bibinfo {author} {\bibfnamefont {L.}~\bibnamefont {Zeng}}, \bibinfo {author}
  {\bibfnamefont {T.}~\bibnamefont {Greibe}}, \bibinfo {author} {\bibfnamefont
  {H.}~\bibnamefont {Pettersson}}, \bibinfo {author} {\bibfnamefont
  {S.}~\bibnamefont {Gustafsson}}, \bibinfo {author} {\bibfnamefont
  {P.}~\bibnamefont {Delsing}}, \ and\ \bibinfo {author} {\bibfnamefont
  {E.}~\bibnamefont {Olsson}},\ }\href@noop {} {\bibfield  {journal} {\bibinfo
  {journal} {SpringerPlus}\ }\textbf {\bibinfo {volume} {5}} (\bibinfo {year}
  {2016})},\ \bibinfo {note} {1067},\ \Eprint
  {http://arxiv.org/abs/https://springerplus.springeropen.com/articles/10.1186/s40064-016-2418-8}
  {https://springerplus.springeropen.com/articles/10.1186/s40064-016-2418-8}
  \BibitemShut {NoStop}%
\bibitem [{\citenamefont {Dimos}\ \emph {et~al.}(1988)\citenamefont {Dimos},
  \citenamefont {Chaudhari}, \citenamefont {Mannhart},\ and\ \citenamefont
  {LeGoues}}]{dimos1988orientation}%
  \BibitemOpen
  \bibfield  {author} {\bibinfo {author} {\bibfnamefont {D.}~\bibnamefont
  {Dimos}}, \bibinfo {author} {\bibfnamefont {P.}~\bibnamefont {Chaudhari}},
  \bibinfo {author} {\bibfnamefont {J.}~\bibnamefont {Mannhart}}, \ and\
  \bibinfo {author} {\bibfnamefont {F.~K.}\ \bibnamefont {LeGoues}},\
  }\href@noop {} {\bibfield  {journal} {\bibinfo  {journal} {Phys. Rev. Lett.}\
  }\textbf {\bibinfo {volume} {61}},\ \bibinfo {pages} {219} (\bibinfo {year}
  {1988})}\BibitemShut {NoStop}%
\bibitem [{\citenamefont {Hilgenkamp}\ and\ \citenamefont
  {Mannhart}(2002)}]{hilgenkamp2002grain}%
  \BibitemOpen
  \bibfield  {author} {\bibinfo {author} {\bibfnamefont {H.}~\bibnamefont
  {Hilgenkamp}}\ and\ \bibinfo {author} {\bibfnamefont {J.}~\bibnamefont
  {Mannhart}},\ }\href@noop {} {\bibfield  {journal} {\bibinfo  {journal} {Rev.
  Mod. Phys.}\ }\textbf {\bibinfo {volume} {74}},\ \bibinfo {pages} {485}
  (\bibinfo {year} {2002})}\BibitemShut {NoStop}%
\bibitem [{\citenamefont {Graser}\ \emph {et~al.}(2010)\citenamefont {Graser},
  \citenamefont {Hirschfeld}, \citenamefont {Kopp}, \citenamefont {Gutser},
  \citenamefont {Andersen},\ and\ \citenamefont {Mannhart}}]{graser2010grain}%
  \BibitemOpen
  \bibfield  {author} {\bibinfo {author} {\bibfnamefont {S.}~\bibnamefont
  {Graser}}, \bibinfo {author} {\bibfnamefont {P.~J.}\ \bibnamefont
  {Hirschfeld}}, \bibinfo {author} {\bibfnamefont {T.}~\bibnamefont {Kopp}},
  \bibinfo {author} {\bibfnamefont {R.}~\bibnamefont {Gutser}}, \bibinfo
  {author} {\bibfnamefont {B.~M.}\ \bibnamefont {Andersen}}, \ and\ \bibinfo
  {author} {\bibfnamefont {J.}~\bibnamefont {Mannhart}},\ }\href@noop {}
  {\bibfield  {journal} {\bibinfo  {journal} {Nature Physics}\ }\textbf
  {\bibinfo {volume} {6}},\ \bibinfo {pages} {609} (\bibinfo {year}
  {2010})}\BibitemShut {NoStop}%
\bibitem [{\citenamefont {Wolf}\ \emph {et~al.}(2012)\citenamefont {Wolf},
  \citenamefont {Graser}, \citenamefont {Loder},\ and\ \citenamefont
  {Kopp}}]{wolf2012supercurrent}%
  \BibitemOpen
  \bibfield  {author} {\bibinfo {author} {\bibfnamefont {F.~A.}\ \bibnamefont
  {Wolf}}, \bibinfo {author} {\bibfnamefont {S.}~\bibnamefont {Graser}},
  \bibinfo {author} {\bibfnamefont {F.}~\bibnamefont {Loder}}, \ and\ \bibinfo
  {author} {\bibfnamefont {T.}~\bibnamefont {Kopp}},\ }\href@noop {} {\bibfield
   {journal} {\bibinfo  {journal} {Phys. Rev. Lett.}\ }\textbf {\bibinfo
  {volume} {108}},\ \bibinfo {pages} {117002} (\bibinfo {year}
  {2012})}\BibitemShut {NoStop}%
\bibitem [{\citenamefont {Furusaki}\ and\ \citenamefont
  {Tsukada}(1991)}]{furusaki1991dc}%
  \BibitemOpen
  \bibfield  {author} {\bibinfo {author} {\bibfnamefont {A.}~\bibnamefont
  {Furusaki}}\ and\ \bibinfo {author} {\bibfnamefont {M.}~\bibnamefont
  {Tsukada}},\ }\href@noop {} {\bibfield  {journal} {\bibinfo  {journal} {Solid
  State Communications}\ }\textbf {\bibinfo {volume} {78}},\ \bibinfo {pages}
  {299} (\bibinfo {year} {1991})}\BibitemShut {NoStop}%
\bibitem [{\citenamefont {Sols}\ and\ \citenamefont
  {Ferrer}(1994)}]{sols1994crossover}%
  \BibitemOpen
  \bibfield  {author} {\bibinfo {author} {\bibfnamefont {F.}~\bibnamefont
  {Sols}}\ and\ \bibinfo {author} {\bibfnamefont {J.}~\bibnamefont {Ferrer}},\
  }\href@noop {} {\bibfield  {journal} {\bibinfo  {journal} {Phys. Rev. B}\
  }\textbf {\bibinfo {volume} {49}},\ \bibinfo {pages} {15913} (\bibinfo {year}
  {1994})}\BibitemShut {NoStop}%
\bibitem [{\citenamefont {Blonder}\ \emph {et~al.}(1982)\citenamefont
  {Blonder}, \citenamefont {Tinkham},\ and\ \citenamefont
  {Klapwijk}}]{blonder1982transition}%
  \BibitemOpen
  \bibfield  {author} {\bibinfo {author} {\bibfnamefont {G.~E.}\ \bibnamefont
  {Blonder}}, \bibinfo {author} {\bibfnamefont {M.}~\bibnamefont {Tinkham}}, \
  and\ \bibinfo {author} {\bibfnamefont {T.~M.}\ \bibnamefont {Klapwijk}},\
  }\href@noop {} {\bibfield  {journal} {\bibinfo  {journal} {Phys. Rev. B}\
  }\textbf {\bibinfo {volume} {25}},\ \bibinfo {pages} {4515} (\bibinfo {year}
  {1982})}\BibitemShut {NoStop}%
\bibitem [{\citenamefont {Furusaki}(1994)}]{furusaki1994dc}%
  \BibitemOpen
  \bibfield  {author} {\bibinfo {author} {\bibfnamefont {A.}~\bibnamefont
  {Furusaki}},\ }\href@noop {} {\bibfield  {journal} {\bibinfo  {journal}
  {Physica B: Condensed Matter}\ }\textbf {\bibinfo {volume} {203}},\ \bibinfo
  {pages} {214} (\bibinfo {year} {1994})}\BibitemShut {NoStop}%
\bibitem [{\citenamefont {Asano}(2001)}]{asano2001numerical}%
  \BibitemOpen
  \bibfield  {author} {\bibinfo {author} {\bibfnamefont {Y.}~\bibnamefont
  {Asano}},\ }\href@noop {} {\bibfield  {journal} {\bibinfo  {journal} {Phys.
  Rev. B}\ }\textbf {\bibinfo {volume} {63}},\ \bibinfo {pages} {052512}
  (\bibinfo {year} {2001})}\BibitemShut {NoStop}%
\bibitem [{\citenamefont {Nikoli{\'c}}\ \emph {et~al.}(2002)\citenamefont
  {Nikoli{\'c}}, \citenamefont {Freericks},\ and\ \citenamefont
  {Miller}}]{nikolic2002equilibrium}%
  \BibitemOpen
  \bibfield  {author} {\bibinfo {author} {\bibfnamefont {B.~K.}\ \bibnamefont
  {Nikoli{\'c}}}, \bibinfo {author} {\bibfnamefont {J.~K.}\ \bibnamefont
  {Freericks}}, \ and\ \bibinfo {author} {\bibfnamefont {P.}~\bibnamefont
  {Miller}},\ }\href@noop {} {\bibfield  {journal} {\bibinfo  {journal} {Phys.
  Rev. B}\ }\textbf {\bibinfo {volume} {65}},\ \bibinfo {pages} {064529}
  (\bibinfo {year} {2002})}\BibitemShut {NoStop}%
\bibitem [{\citenamefont {Andersen}\ \emph {et~al.}(2005)\citenamefont
  {Andersen}, \citenamefont {Bobkova}, \citenamefont {Hirschfeld},\ and\
  \citenamefont {Barash}}]{andersen2005bound}%
  \BibitemOpen
  \bibfield  {author} {\bibinfo {author} {\bibfnamefont {B.~M.}\ \bibnamefont
  {Andersen}}, \bibinfo {author} {\bibfnamefont {I.~V.}\ \bibnamefont
  {Bobkova}}, \bibinfo {author} {\bibfnamefont {P.~J.}\ \bibnamefont
  {Hirschfeld}}, \ and\ \bibinfo {author} {\bibfnamefont {Y.~S.}\ \bibnamefont
  {Barash}},\ }\href@noop {} {\bibfield  {journal} {\bibinfo  {journal} {Phys.
  Rev. B}\ }\textbf {\bibinfo {volume} {72}},\ \bibinfo {pages} {184510}
  (\bibinfo {year} {2005})}\BibitemShut {NoStop}%
\bibitem [{\citenamefont {Andersen}\ \emph {et~al.}(2006)\citenamefont
  {Andersen}, \citenamefont {Bobkova}, \citenamefont {Hirschfeld},\ and\
  \citenamefont {Barash}}]{andersen20060}%
  \BibitemOpen
  \bibfield  {author} {\bibinfo {author} {\bibfnamefont {B.~M.}\ \bibnamefont
  {Andersen}}, \bibinfo {author} {\bibfnamefont {I.~V.}\ \bibnamefont
  {Bobkova}}, \bibinfo {author} {\bibfnamefont {P.~J.}\ \bibnamefont
  {Hirschfeld}}, \ and\ \bibinfo {author} {\bibfnamefont {Y.~S.}\ \bibnamefont
  {Barash}},\ }\href@noop {} {\bibfield  {journal} {\bibinfo  {journal} {Phys.
  Rev. Lett.}\ }\textbf {\bibinfo {volume} {96}},\ \bibinfo {pages} {117005}
  (\bibinfo {year} {2006})}\BibitemShut {NoStop}%
\bibitem [{\citenamefont {Covaci}\ and\ \citenamefont
  {Marsiglio}(2006)}]{covaci2006proximity}%
  \BibitemOpen
  \bibfield  {author} {\bibinfo {author} {\bibfnamefont {L.}~\bibnamefont
  {Covaci}}\ and\ \bibinfo {author} {\bibfnamefont {F.}~\bibnamefont
  {Marsiglio}},\ }\href@noop {} {\bibfield  {journal} {\bibinfo  {journal}
  {Phys. Rev. B}\ }\textbf {\bibinfo {volume} {73}},\ \bibinfo {pages} {014503}
  (\bibinfo {year} {2006})}\BibitemShut {NoStop}%
\bibitem [{\citenamefont {Andersen}\ \emph {et~al.}(2008)\citenamefont
  {Andersen}, \citenamefont {Barash}, \citenamefont {Graser},\ and\
  \citenamefont {Hirschfeld}}]{andersen2008josephson}%
  \BibitemOpen
  \bibfield  {author} {\bibinfo {author} {\bibfnamefont {B.~M.}\ \bibnamefont
  {Andersen}}, \bibinfo {author} {\bibfnamefont {Y.~S.}\ \bibnamefont
  {Barash}}, \bibinfo {author} {\bibfnamefont {S.}~\bibnamefont {Graser}}, \
  and\ \bibinfo {author} {\bibfnamefont {P.~J.}\ \bibnamefont {Hirschfeld}},\
  }\href@noop {} {\bibfield  {journal} {\bibinfo  {journal} {Phys. Rev. B}\
  }\textbf {\bibinfo {volume} {77}},\ \bibinfo {pages} {054501} (\bibinfo
  {year} {2008})}\BibitemShut {NoStop}%
\bibitem [{\citenamefont {Black-Schaffer}\ and\ \citenamefont
  {Doniach}(2008)}]{black2008self}%
  \BibitemOpen
  \bibfield  {author} {\bibinfo {author} {\bibfnamefont {A.~M.}\ \bibnamefont
  {Black-Schaffer}}\ and\ \bibinfo {author} {\bibfnamefont {S.}~\bibnamefont
  {Doniach}},\ }\href@noop {} {\bibfield  {journal} {\bibinfo  {journal} {Phys.
  Rev. B}\ }\textbf {\bibinfo {volume} {78}},\ \bibinfo {pages} {024504}
  (\bibinfo {year} {2008})}\BibitemShut {NoStop}%
\bibitem [{\citenamefont {Plimpton}(1995)}]{Plimpton1995}%
  \BibitemOpen
  \bibfield  {author} {\bibinfo {author} {\bibfnamefont {S.}~\bibnamefont
  {Plimpton}},\ }\href {\doibase https://doi.org/10.1006/jcph.1995.1039}
  {\bibfield  {journal} {\bibinfo  {journal} {Journal of Computational
  Physics}\ }\textbf {\bibinfo {volume} {117}},\ \bibinfo {pages} {1} (\bibinfo
  {year} {1995})}\BibitemShut {NoStop}%
\bibitem [{\citenamefont {Garg}\ \emph
  {et~al.}(2020{\natexlab{b}})\citenamefont {Garg}, \citenamefont {Muhich},
  \citenamefont {Cooley}, \citenamefont {Bieler},\ and\ \citenamefont
  {Solanki}}]{Garg2020}%
  \BibitemOpen
  \bibfield  {author} {\bibinfo {author} {\bibfnamefont {P.}~\bibnamefont
  {Garg}}, \bibinfo {author} {\bibfnamefont {C.}~\bibnamefont {Muhich}},
  \bibinfo {author} {\bibfnamefont {L.~D.}\ \bibnamefont {Cooley}}, \bibinfo
  {author} {\bibfnamefont {T.~R.}\ \bibnamefont {Bieler}}, \ and\ \bibinfo
  {author} {\bibfnamefont {K.~N.}\ \bibnamefont {Solanki}},\ }\href {\doibase
  10.1103/PhysRevB.101.184102} {\bibfield  {journal} {\bibinfo  {journal}
  {Phys. Rev. B}\ }\textbf {\bibinfo {volume} {101}},\ \bibinfo {pages}
  {184102} (\bibinfo {year} {2020}{\natexlab{b}})}\BibitemShut {NoStop}%
\bibitem [{\citenamefont {Fellinger}\ \emph {et~al.}(2010)\citenamefont
  {Fellinger}, \citenamefont {Park},\ and\ \citenamefont
  {Wilkins}}]{Fellinger2010}%
  \BibitemOpen
  \bibfield  {author} {\bibinfo {author} {\bibfnamefont {M.~R.}\ \bibnamefont
  {Fellinger}}, \bibinfo {author} {\bibfnamefont {H.}~\bibnamefont {Park}}, \
  and\ \bibinfo {author} {\bibfnamefont {J.~W.}\ \bibnamefont {Wilkins}},\
  }\href {\doibase 10.1103/PhysRevB.81.144119} {\bibfield  {journal} {\bibinfo
  {journal} {Phys. Rev. B}\ }\textbf {\bibinfo {volume} {81}},\ \bibinfo
  {pages} {144119} (\bibinfo {year} {2010})}\BibitemShut {NoStop}%
\bibitem [{\citenamefont {Rittner}\ and\ \citenamefont
  {Seidman}(1996)}]{Rittner1996}%
  \BibitemOpen
  \bibfield  {author} {\bibinfo {author} {\bibfnamefont {J.~D.}\ \bibnamefont
  {Rittner}}\ and\ \bibinfo {author} {\bibfnamefont {D.~N.}\ \bibnamefont
  {Seidman}},\ }\href {\doibase 10.1103/PhysRevB.54.6999} {\bibfield  {journal}
  {\bibinfo  {journal} {Phys. Rev. B}\ }\textbf {\bibinfo {volume} {54}},\
  \bibinfo {pages} {6999} (\bibinfo {year} {1996})}\BibitemShut {NoStop}%
\bibitem [{\citenamefont {Rajagopalan}\ \emph {et~al.}(2017)\citenamefont
  {Rajagopalan}, \citenamefont {Adlakha}, \citenamefont {Tschopp},\ and\
  \citenamefont {Solanki}}]{Rajagopalan2017}%
  \BibitemOpen
  \bibfield  {author} {\bibinfo {author} {\bibfnamefont {M.}~\bibnamefont
  {Rajagopalan}}, \bibinfo {author} {\bibfnamefont {I.}~\bibnamefont
  {Adlakha}}, \bibinfo {author} {\bibfnamefont {M.~A.}\ \bibnamefont
  {Tschopp}}, \ and\ \bibinfo {author} {\bibfnamefont {K.~N.}\ \bibnamefont
  {Solanki}},\ }\href {\doibase 10.1007/s11837-017-2386-7} {\bibfield
  {journal} {\bibinfo  {journal} {JOM}\ }\textbf {\bibinfo {volume} {69}},\
  \bibinfo {pages} {1398} (\bibinfo {year} {2017})}\BibitemShut {NoStop}%
\bibitem [{\citenamefont {Adlakha}\ \emph {et~al.}(2014)\citenamefont
  {Adlakha}, \citenamefont {Bhatia}, \citenamefont {Tschopp},\ and\
  \citenamefont {Solanki}}]{Adlakha2014}%
  \BibitemOpen
  \bibfield  {author} {\bibinfo {author} {\bibfnamefont {I.}~\bibnamefont
  {Adlakha}}, \bibinfo {author} {\bibfnamefont {M.}~\bibnamefont {Bhatia}},
  \bibinfo {author} {\bibfnamefont {M.}~\bibnamefont {Tschopp}}, \ and\
  \bibinfo {author} {\bibfnamefont {K.}~\bibnamefont {Solanki}},\ }\href
  {\doibase 10.1080/14786435.2014.961585} {\bibfield  {journal} {\bibinfo
  {journal} {Philosophical Magazine}\ }\textbf {\bibinfo {volume} {94}},\
  \bibinfo {pages} {3445} (\bibinfo {year} {2014})},\ \Eprint
  {http://arxiv.org/abs/https://doi.org/10.1080/14786435.2014.961585}
  {https://doi.org/10.1080/14786435.2014.961585} \BibitemShut {NoStop}%
\bibitem [{\citenamefont {Rajagopalan}\ \emph {et~al.}(2014)\citenamefont
  {Rajagopalan}, \citenamefont {Bhatia}, \citenamefont {Tschopp}, \citenamefont
  {Srolovitz},\ and\ \citenamefont {Solanki}}]{Rajagopalan2014}%
  \BibitemOpen
  \bibfield  {author} {\bibinfo {author} {\bibfnamefont {M.}~\bibnamefont
  {Rajagopalan}}, \bibinfo {author} {\bibfnamefont {M.}~\bibnamefont {Bhatia}},
  \bibinfo {author} {\bibfnamefont {M.}~\bibnamefont {Tschopp}}, \bibinfo
  {author} {\bibfnamefont {D.}~\bibnamefont {Srolovitz}}, \ and\ \bibinfo
  {author} {\bibfnamefont {K.}~\bibnamefont {Solanki}},\ }\href {\doibase
  https://doi.org/10.1016/j.actamat.2014.04.011} {\bibfield  {journal}
  {\bibinfo  {journal} {Acta Materialia}\ }\textbf {\bibinfo {volume} {73}},\
  \bibinfo {pages} {312} (\bibinfo {year} {2014})}\BibitemShut {NoStop}%
\bibitem [{\citenamefont {Solanki}\ \emph {et~al.}(2013)\citenamefont
  {Solanki}, \citenamefont {Tschopp}, \citenamefont {Bhatia},\ and\
  \citenamefont {Rhodes}}]{Solanki2013}%
  \BibitemOpen
  \bibfield  {author} {\bibinfo {author} {\bibfnamefont {K.~N.}\ \bibnamefont
  {Solanki}}, \bibinfo {author} {\bibfnamefont {M.~A.}\ \bibnamefont
  {Tschopp}}, \bibinfo {author} {\bibfnamefont {M.~A.}\ \bibnamefont {Bhatia}},
  \ and\ \bibinfo {author} {\bibfnamefont {N.~R.}\ \bibnamefont {Rhodes}},\
  }\href {\doibase 10.1007/s11661-012-1430-z} {\bibfield  {journal} {\bibinfo
  {journal} {Metallurgical and Materials Transactions A}\ }\textbf {\bibinfo
  {volume} {44}},\ \bibinfo {pages} {1365} (\bibinfo {year}
  {2013})}\BibitemShut {NoStop}%
\bibitem [{\citenamefont {Tschopp}\ \emph {et~al.}(2012)\citenamefont
  {Tschopp}, \citenamefont {Solanki}, \citenamefont {Gao}, \citenamefont {Sun},
  \citenamefont {Khaleel},\ and\ \citenamefont {Horstemeyer}}]{Tschopp2012}%
  \BibitemOpen
  \bibfield  {author} {\bibinfo {author} {\bibfnamefont {M.~A.}\ \bibnamefont
  {Tschopp}}, \bibinfo {author} {\bibfnamefont {K.~N.}\ \bibnamefont
  {Solanki}}, \bibinfo {author} {\bibfnamefont {F.}~\bibnamefont {Gao}},
  \bibinfo {author} {\bibfnamefont {X.}~\bibnamefont {Sun}}, \bibinfo {author}
  {\bibfnamefont {M.~A.}\ \bibnamefont {Khaleel}}, \ and\ \bibinfo {author}
  {\bibfnamefont {M.~F.}\ \bibnamefont {Horstemeyer}},\ }\href {\doibase
  10.1103/PhysRevB.85.064108} {\bibfield  {journal} {\bibinfo  {journal} {Phys.
  Rev. B}\ }\textbf {\bibinfo {volume} {85}},\ \bibinfo {pages} {064108}
  (\bibinfo {year} {2012})}\BibitemShut {NoStop}%
\bibitem [{\citenamefont {Bruin}\ \emph {et~al.}(2013)\citenamefont {Bruin},
  \citenamefont {Sakai}, \citenamefont {Perry},\ and\ \citenamefont
  {Mackenzie}}]{Bruin2013}%
  \BibitemOpen
  \bibfield  {author} {\bibinfo {author} {\bibfnamefont {J.~A.~N.}\
  \bibnamefont {Bruin}}, \bibinfo {author} {\bibfnamefont {H.}~\bibnamefont
  {Sakai}}, \bibinfo {author} {\bibfnamefont {R.~S.}\ \bibnamefont {Perry}}, \
  and\ \bibinfo {author} {\bibfnamefont {A.~P.}\ \bibnamefont {Mackenzie}},\
  }\href {\doibase 10.1126/science.1227612} {\bibfield  {journal} {\bibinfo
  {journal} {Science}\ }\textbf {\bibinfo {volume} {339}},\ \bibinfo {pages}
  {804} (\bibinfo {year} {2013})}\BibitemShut {NoStop}%
\bibitem [{\citenamefont {Koepernik}\ and\ \citenamefont
  {Eschrig}(1999)}]{KoepernikFPLO}%
  \BibitemOpen
  \bibfield  {author} {\bibinfo {author} {\bibfnamefont {K.}~\bibnamefont
  {Koepernik}}\ and\ \bibinfo {author} {\bibfnamefont {H.}~\bibnamefont
  {Eschrig}},\ }\href {\doibase 10.1103/PhysRevB.59.1743} {\bibfield  {journal}
  {\bibinfo  {journal} {Phys. Rev. B}\ }\textbf {\bibinfo {volume} {59}},\
  \bibinfo {pages} {1743} (\bibinfo {year} {1999})}\BibitemShut {NoStop}%
\bibitem [{Note1()}]{Note1}%
  \BibitemOpen
  \bibinfo {note} {Obtained from a DFT calculation using the lattice constant
  of 2.87565467\r A{} in the primitive setting and local orbital basis in the
  LDSA approximation using the code based on Ref. 40; Y. Quan, private
  communication.}\BibitemShut {Stop}%
\bibitem [{\citenamefont {Hahn}\ \emph {et~al.}(1998)\citenamefont {Hahn},
  \citenamefont {Hofmann}, \citenamefont {Krause},\ and\ \citenamefont
  {Seidel}}]{Hahn98}%
  \BibitemOpen
  \bibfield  {author} {\bibinfo {author} {\bibfnamefont {A.}~\bibnamefont
  {Hahn}}, \bibinfo {author} {\bibfnamefont {S.}~\bibnamefont {Hofmann}},
  \bibinfo {author} {\bibfnamefont {A.}~\bibnamefont {Krause}}, \ and\ \bibinfo
  {author} {\bibfnamefont {P.}~\bibnamefont {Seidel}},\ }\href {\doibase
  https://doi.org/10.1016/S0921-4534(97)01826-1} {\bibfield  {journal}
  {\bibinfo  {journal} {Physica C: Superconductivity}\ }\textbf {\bibinfo
  {volume} {296}},\ \bibinfo {pages} {103} (\bibinfo {year}
  {1998})}\BibitemShut {NoStop}%
\bibitem [{\citenamefont {Rogachev}\ and\ \citenamefont
  {Bezryadin}(2003)}]{rogachev2003superconducting}%
  \BibitemOpen
  \bibfield  {author} {\bibinfo {author} {\bibfnamefont {A.}~\bibnamefont
  {Rogachev}}\ and\ \bibinfo {author} {\bibfnamefont {A.}~\bibnamefont
  {Bezryadin}},\ }\href@noop {} {\bibfield  {journal} {\bibinfo  {journal}
  {Applied Physics Letters}\ }\textbf {\bibinfo {volume} {83}},\ \bibinfo
  {pages} {512} (\bibinfo {year} {2003})}\BibitemShut {NoStop}%
\end{thebibliography}%

\end{document}